\documentclass[article,12pt]{article}
 \usepackage{amsmath}
\usepackage{shadethm}
\usepackage{dsfont}
\usepackage{float}
\usepackage{fancyhdr}
\usepackage{array,multirow}
\usepackage{amssymb}
\usepackage{diagbox}
\usepackage{graphicx}
\usepackage{color}
\usepackage{natbib}
\usepackage{algpseudocode}
\usepackage[dvipsnames]{xcolor}
\DeclareMathOperator*{\argmax}{arg\,max}  
\DeclareMathOperator*{\argmin}{arg\,min}
\usepackage{xcolor}
\usepackage{tcolorbox} 
\usepackage{bm}
\usepackage[utf8]{inputenc}
\usepackage{amsfonts}
\usepackage{subcaption}

\usepackage{verbatim}
\usepackage{lscape}
\usepackage[titles,subfigure]{tocloft}
\usepackage[colorlinks,bookmarksopen,bookmarksnumbered,citecolor=red,urlcolor=red]{hyperref}
\usepackage{setspace}
\usepackage{titlesec}
\usepackage{enumitem}
\usepackage{longtable} 
\usepackage[utf8]{inputenc}   
\usepackage[T1]{fontenc}

\usepackage[figuresright]{rotating}

\definecolor{chartreuse}{rgb}{0.5, 1.0, 0.0}
\definecolor{applegreen}{rgb}{0.55, 0.9, 0.0}
\definecolor{green(ryb)}{rgb}{0.4, 0.6, 0.3}
\definecolor{green(light)}{rgb}{0.4, 0.9, 0.3}

\DeclareMathOperator*{\pen}{pen}
\DeclareMathOperator*{\sgn}{sgn}
\DeclareMathOperator*{\MLE}{MLE}
\usepackage{graphicx}
\graphicspath{/Figures}
\usepackage[normalem]{ulem}

\usepackage[a4paper,top=2.5cm,bottom=5cm,left=3.6cm,right=2.4cm]{geometry}
\usepackage{xr} 
\usepackage{hyperref} 
\usepackage{authblk}
\title{A regularized multi-state model for covariate selection with interval-censored survival data}

\author[1]{Ariane Bercu}
\author[2]{Agathe Guilloux}
\author[1]{C\'ecile Proust-Lima$^\dagger$}
\author[1]{H\'el\`ene Jacqmin-Gadda$^\dagger$}

\affil[1]{\small Univ. Bordeaux, INSERM, BPH, U1219, F-33000 Bordeaux, France}
\affil[2]{\small Inria Paris F-75015, INSERM, CRC, Univ. Sorbonne, F-75006 Paris, France}
\affil[ ]{\small $^\dagger$Co-last authors}

\affil[ ]{\texttt{\small ariane.bercu@u-bordeaux.fr, agathe.guilloux@inria.fr, 
cecile.proust-lima@u-bordeaux.fr, helene.jacqmin-gadda@u-bordeaux.fr}}

\begin{document}

\maketitle
\vspace{-2ex}
\begin{abstract}

In population-based cohorts, disease diagnoses are typically censored by intervals as made during scheduled follow-up visits. The exact disease onset time is thus unknown, and in the presence of semi-competing risk of death, subjects may also die in between two visits before any diagnosis can be made. Illness-death models can be used to handle uncertainty about illness timing and the possible absence of diagnosis due to death. However, they are so far limited in the number of covariates. We developed a regularized estimation procedure for illness-death models with interval-censored illness diagnosis that performs variable selection in the case of high-dimensional predictors. We considered a proximal gradient hybrid algorithm maximizing the regularized likelihood with an elastic-net penalty. The algorithm simultaneously estimates the regression parameters of the three transitions under proportional transition intensities with transition-specific penalty parameters determined in an outer gridsearch. The algorithm, implemented in the R package HIDeM, shows high performances in predicting illness probability, as well as correct selection of transition-specific risk factors across different simulation scenarios. In comparison, the cause-specific competing risk model neglecting interval-censoring systematically showed worse predictive ability and tended to select irrelevant illness predictors, originally associated with death.
Applied to the population-based cohort Three-City, the method identified predictors of clinical dementia onset among a large set of brain imaging, cognitive and clinical markers. \\
\vspace{-2ex}

\noindent  {\bf Keywords: }Interval censoring; Multi-state model; Semi-competing risk;  Survival Analysis; Variable Selection. 

\end{abstract}

\newpage
\section{Introduction} 
 Interval-censored times to disease onset are common in population-based cohort studies, when the disease diagnosis is only assessed during scheduled visits, making it impossible to retrospectively determine the exact date of disease onset \citep{chen_interval-censored_2012}. When a subject is diagnosed, the time to disease onset is interval-censored between the last visit at which he was illness-free and the visit at which the diagnosis occurs. If the subject dies after the disease onset but before the diagnosis can be established at a scheduled visit, the disease remains undiagnosed, leading to an underestimation of the disease incidence \citep{leffondre_interval-censored_2013}. This additional issue comes from the fact that death and the disease are semi-competing events: while death precludes the disease onset, it may still occur after the disease has occurred.

These issues arise in most longitudinal studies on dementia. For example, the Three-City Study (3C Study) \citep{the_3c_study_group_vascular_2003} is a French population-based longitudinal study investigating the relationship between vascular diseases and dementia in individuals aged 65 years and older. In the 3C study, dementia diagnosis is assessed only at planned visits every 2 or 3 years, allowing for the possibility that subjects develop dementia and die between two visits. The risk of bias is particularly high in such studies because the mortality rate is high in the elderly population, it increases for subjects with dementia and because many risk factors are common for dementia and death \citep{liang_mortality_2021}. In such a situation, a naive approach using a standard survival model can be applied by imputing the middle of the interval for subjects diagnosed with dementia and considering the other subjects as censored at their last visit. Due to the induced informative censoring and uncertainty on time of disease, \cite{leffondre_interval-censored_2013} have shown that both the estimated incidence of dementia and the estimated associations between risk factors and dementia are biased. Standard models for competing risks \citep{fine_semi-competing_2001,andersen_competing_2012} are also not suitable because they require the knowledge of the type and the exact time of the first event, which is unknown 
for diagnosed subjects and non-diagnosed deceased subjects.

\cite{joly_penalized_2002} developed an Illness-Death regression Model (IDM) to handle interval-censored data in which all individuals enter in the healthy state, may develop the disease, and may die either with or without the disease. The estimation procedure accounts for the uncertainty about the disease onset time and the possibly non-diagnosed disease cases. The IDM has proven its superiority over the semi-competing risk Proportional Hazard Model (PHM) with unspecified (Cox) or Weibull baseline risks \citep{leffondre_interval-censored_2013} in the presence of interval censoring. 
Nonetheless, its practical application remains marginal, partly because its implementation is not suited to handle a large number of explanatory variables.

Indeed, the rapid advancement and wider accessibility of technology have allowed population-based studies to collect progressively richer information (including brain volumes from imaging data, fluid biomarkers, detailed cognitive and functional evaluation), which in turn complicates the identification of risk factors. The need to select relevant covariates has driven the development of variable selection methods in survival analysis. Variable selection methods based on regularization have been widely extended to survival analysis to identify risk factors and improve prediction accuracy. The most popular one is the Least Absolute Shrinkage and Selection Operator (LASSO) proposed by \cite{tibshirani_regression_1996}. \cite{zou_regularization_2005} defined the elastic net penalty as a combination of LASSO and ridge penalties, allowing for variable selection even in the presence of correlated covariates. To our knowledge, only a few publications are available on regularization with interval-censored time. \cite{li_adaptive_2020} developed an adaptive LASSO in Cox's PHM dealing with uncertain time to disease onset. \cite{zhao_simultaneous_2020} implemented a simultaneous estimation and variable selection for Cox's PHM using Broken Adaptive Ridge (BAR) with uncertainty about the time to disease onset. \cite{feng_variable_2023} proposed various penalties such as LASSO, adaptive LASSO and BAR under PHM with an uncertain time to disease onset. While these approaches address the uncertainty in the time to disease onset, they disregard the uncertainty in the illness status itself, which is the main issue in the analysis of interval-censored data with semi-competing events.

In this paper, we propose a new regularized estimation procedure that jointly addresses the challenges of high-dimensional covariates, interval censoring and semi-competing risks. Specifically, we develop an efficient algorithm to maximize the regularized likelihood of an illness-death model with interval-censored semi-competing event times, incorporating elastic net penalization to allow variable selection.
Section~\ref{methodo} introduces the regularized estimation framework and describes the optimization strategy. Section~\ref{simu} presents simulation studies that evaluate the performance of the proposed approach and compare it to that of the standard competing risk PHM, which ignores interval censoring. Section~\ref{appli} illustrates the method with an application to dementia research, aiming to identify key predictors of dementia risk using data from the Three-City (3C) population-based study \citep{the_3c_study_group_vascular_2003}. Finally, Section~\ref{discussion} discusses the strengths and limitations of the method and outlines directions for future research.

\section{Methodology} \label{methodo}
\subsection{Notations}
 We denote by $T^{I}_i$ the unobserved time to illness (i.e., disease onset), and by $T^D_i$ the time of death for individual $i$ ($i = 1, \ldots, N$). We assume that $T^{I}_i$ is interval-censored, whereas $T^D_i$ is only subject to right-censoring, as exact dates of death are generally available in population-based studies.
The time-to-event variables and associated indicators collected for individual $i$ are recorded in the vector $D_i = (V_{0i}, L_i, R_i, \delta^{I}_i, T_i, \delta^{D}i)$, where $V_{0i}$ denotes the entry visit time, $L_i$ the time of the last visit at which the individual was observed healthy, $R_i$ the time of the diagnosis visit if the individual was diagnosed with the disease, $T_i$ the minimum between the time of death and the end of follow-up, $\delta^{I}_i$ indicates whether the individual was diagnosed with the disease ($\delta^{I}_i = 1$ if diagnosed, 0 otherwise), and $\delta^{D}_i$ indicates whether the individual died during follow-up ($\delta^{D}_i = 1$ if deceased, 0 otherwise).

\subsection{The Illness-Death Model (IDM)}\label{defIDM}
Multi-state models introduced by \cite{andersen_competing_2002} describe the evolution of individuals through successive health states. The illness-death model is a multi-state model with 3 states where an individual may experience the following transitions: {\em healthy-diseased} (0 $\rightarrow$ 1), {\em diseased-deceased} (1 $\rightarrow$ 2), and {\em healthy-deceased} (0 $\rightarrow$ 2).  These transitions are governed by transition-specific intensities. Assuming proportional transition intensities, the transition intensity from state $h$ to state $l$ at time $t$ for the  individual $i$ is given by
\begin{equation*} \label{eq:2}
\alpha_{hli}(t|\boldsymbol{\theta_{hl}}, \boldsymbol{\beta_{hl}}, \boldsymbol{ Z_{i}}) = \alpha_{0,hl}(t|\boldsymbol{\theta_{hl}})e^{\boldsymbol{\beta_{hl}^\top Z_{i}}}
\end{equation*}
where $\boldsymbol{Z_{i}}$ denotes the vector of explanatory variables for individual $i$, $\boldsymbol{\beta_{hl}}$ are the associated regression coefficients and $\alpha_{0,hl}(t | \boldsymbol{\theta_{hl}})$ the baseline transition intensity for $h \rightarrow l$ at time $t$ parametrized by $\boldsymbol{\theta_{hl}}$. In this work, we consider two parametric baseline intensities: a Weibull baseline transition intensity with 
\begin{equation*}
\alpha_{0,hl}(t|\theta_{1,hl},\theta_{2,hl})=\theta_{1,hl}\theta_{2,hl}(\theta_{2,hl}t)^{\theta_{1,hl}-1}
\end{equation*}
and a more flexible baseline transition intensity defined on a basis of cubic M-splines 
\begin{equation*} 
\alpha_{0,hl}(t|\theta_{1,hl},\ldots, \theta_{K,hl})=\sum_{k=1}^K \theta_{k,hl}M_k(t) 
\end{equation*} 
with $K-2$ the number of nodes ($K\geq$4) and  $M_k(t)$ the $k^{th}$ M-spline function evaluated in $t$. Individual cumulative transition intensity from state $h$ to $l$ up to time $t$ is denoted by $A_{hli}(t|\boldsymbol{\theta_{hl}},\boldsymbol{\beta_{hl}}, \boldsymbol{Z_{i}})$, with $A_{hli}(t|\boldsymbol{\theta_{hl}},\boldsymbol{\beta_{hl}}, \boldsymbol{Z_{i}})=\int_{0}^t\alpha_{hli}(u|\boldsymbol{\theta_{hl}},\boldsymbol{\beta_{hl}}, \boldsymbol{Z_{i}})du$. For brevity, we omit the dependence on parameters and covariates in the following, and simply write $\alpha_{hli}(t)$ and $A_{hli}(t)$.

\subsection{Individual contribution to the likelihood with interval-censoring} 

\begin{figure}[htbp]
    \centering
    \includegraphics[clip, trim=0cm 10cm 0cm 13cm, width=\textwidth]{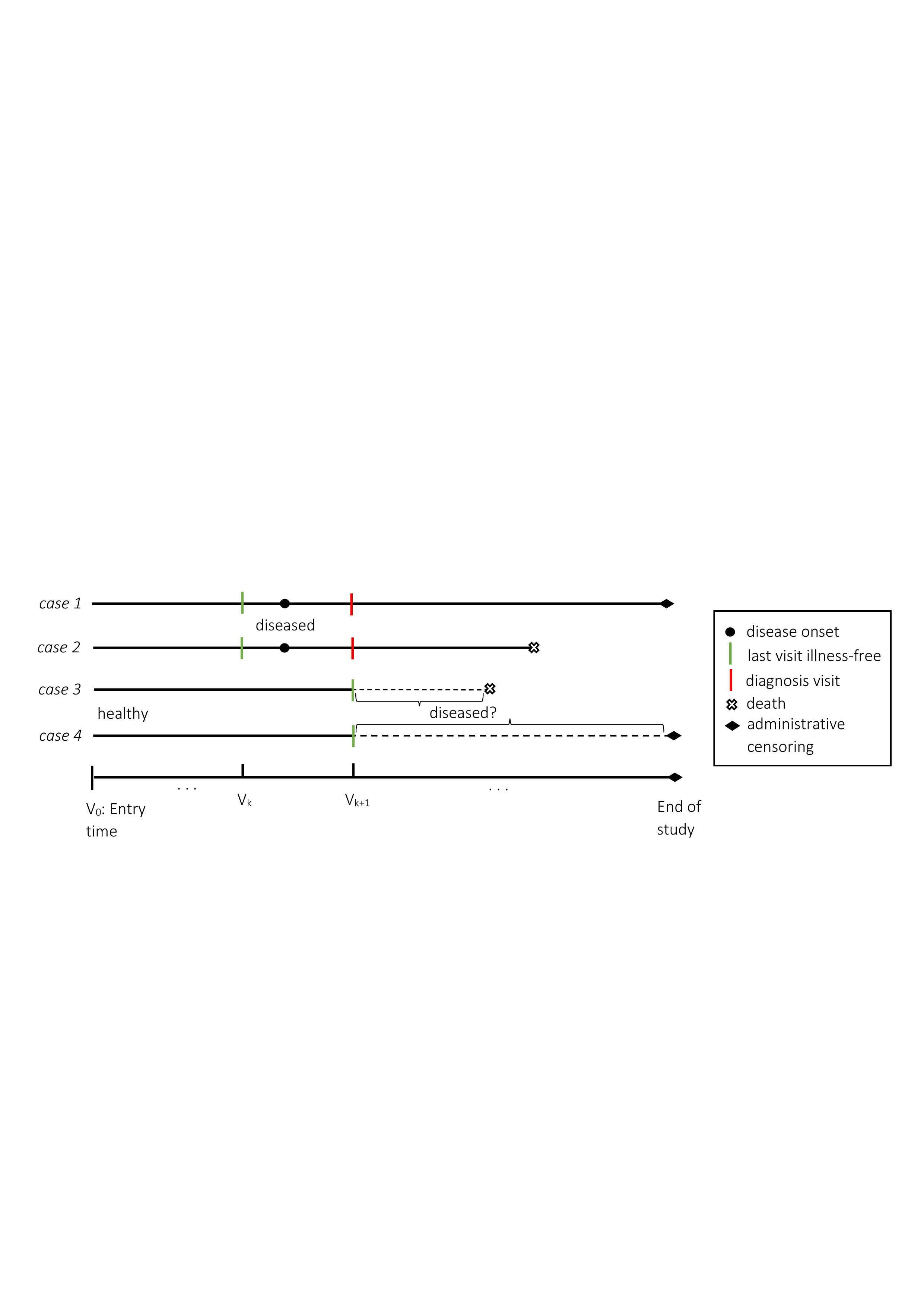}
    \caption{The four follow-up patterns in the presence of interval-censored time to disease onset, and semi-competing risk of death.}
    \label{suivi}
\end{figure}

Based on the observation patterns of population-based studies, schematized in Figure \ref{suivi}, the individual contribution to the IDM likelihood defined in \cite{joly_penalized_2002} varies by case:
\begin{itemize}
    \item  The individual remains healthy until visit $L_{i}$, is diagnosed with the disease at visit time $R_{i}$ and is either censored at time $T_i$ (case 1) or dies at $T_i$ (case 2), 
\begin{align*}
    \mathcal{L}_i =\int_{L_{i}}^{R_{i}}e^{-A_{01i}(u)-A_{02i}(u)}\alpha_{01i}(u)e^{-A_{12i}(T_i) +A_{12i}(u)}du ~\alpha_{12i}(T_i)^{\delta^D_{i}}.
\end{align*}
The integral corresponds to the probability of becoming ill in the interval $[L_i, R_i]$ and remaining alive until time $T_i$. The last term accounts for individuals who died at time $T_i$.
\item The individual remains healthy until visit $L_{i}$ and is  either censored at time $T_i$ (case 3) or dies at $T_i$ (case 4),
\begin{align*}\mathcal{L}_i &=  \displaystyle{ e^{-A_{01i}(T_i)-A_{02i}(T_i)} \alpha_{02i}(T_i)^{\delta^D_{i}}} \\&+ 
     \int_{L_i}^{T_i}e^{-A_{01i}(u)-A_{02i}(u)}\alpha_{01i}(u)e^{-A_{12i}(T_i)+A_{12i}(u)}du ~ \alpha_{12i}(T_i)^{\delta^D_{i}}.\end{align*}
The first term accounts for the possible direct transition from healthy to death, while the second term considers a possible transition through the disease at time $u$ between $L_{i}$ and $T_i$.
\end{itemize}

The overall likelihood is obtained by multiplying the individual likelihood contributions of each individual $\mathcal{L}_i$ with $i=1,\ldots,N$. Thus, the log-likelihood is given by 
\begin{equation*}
    \ell=\log(\mathcal{L}) = \sum_{i=1}^N \log(\mathcal{L}_i).
\end{equation*} 
For studying death and dementia, the most sensible time-scale is age as it is the main risk factor for both events. However  data become left-truncated, as subjects enter the study at different ages and individuals must be alive and illness-free at study entry to be included in the sample \citep{commenges_dynamical_2015}. The individual likelihood contribution has thus to be divided by the probability of staying alive without the disease until the first visit $V_{0i}$,
\begin{equation*}\mathcal{L}_i=\frac{\mathcal{L}_i}{e^{-A_{01i}(V_{0i})-A_{02i}(V_{0i})}}. 
\end{equation*}
Although the methodology and software can account for left truncation, for simplicity, we illustrate our method in the rest of manuscript using time since entry as the time scale (assuming $V_{0i}=0$ for all individuals), so the data are free from left truncation. 

\subsection{Regularized log-likelihood maximization} 

In modern population-based studies, the large number of available covariates — potential risk factors — leads to high-dimensional explanatory vectors $\boldsymbol{Z_{i}}$ for each transition. The number of regression coefficients to estimate becomes large, making regularization essential. We thus developed and implemented a regularized maximum likelihood estimation procedure for the IDM with interval-censored data, using an elastic net penalty.
In Subsection~\ref{regularized-loglik}, we define the elastic net regularized log-likelihood. Subsection ~\ref{innerloop} details our optimization procedure for fixed hyperparameters, and Subsection~\ref{pena_choice} presents the strategy for selecting these hyperparameters.

\subsubsection{Regularized log-likelihood} \label{regularized-loglik}

Maximizing the regularized log-likelihood, denoted $\ell_{pen}$, corresponds to solving the following optimization problem
\begin{align}
    \label{eq:1}
(\boldsymbol{\widehat\theta}(a,\boldsymbol{\lambda}),\boldsymbol{\widehat\beta}(a,\boldsymbol{\lambda})) &=  \underset{\boldsymbol{\theta},\boldsymbol{\beta}}{\argmax}\{\ell_{\pen}(\boldsymbol{\theta}(a,\boldsymbol{\lambda}),\boldsymbol{\beta}(a,\boldsymbol{\lambda})) \}  \\&= \underset{\boldsymbol{\theta},\boldsymbol{\beta}}{\argmax}\{\ell(\boldsymbol{\theta},\boldsymbol{\beta}) - \pen(\boldsymbol{\beta},a,\boldsymbol{\lambda})\} \nonumber
\end{align} 

where $\ell(\boldsymbol{\theta}, \boldsymbol{\beta})$ is the (unpenalized) log-likelihood evaluated at $(\boldsymbol{\theta},\boldsymbol{\beta})$, $\pen(\boldsymbol{\beta}, a, \boldsymbol{\lambda})$ is the elastic-net penalty, $\boldsymbol{\theta = (\theta_{01}, \theta_{02}, \theta_{12})}$ are the baseline intensity parameters for each transition, $\boldsymbol{\beta = (\beta_{01}, \beta_{02}, \beta_{12})}$ are the regression coefficients associated with each transition, and $\boldsymbol{\lambda }= (\lambda_{01}, \lambda_{02}, \lambda_{12})$ are the transition-specific penalty weights. The total penalty function decomposes over the transitions as:
$$
\pen(\boldsymbol{\beta}, a, \boldsymbol{\lambda}) = \pen(\boldsymbol{\beta_{01}}, a, \lambda_{01}) + \pen(\boldsymbol{\beta_{02}}, a, \lambda_{02}) + \pen(\boldsymbol{\beta_{12}}, a, \lambda_{12}),
$$
with each transition-specific penalty defined by:
$$
\pen(\boldsymbol{\beta_{hl}}, a, \lambda_{hl}) = \lambda_{hl} \left[ a \| \boldsymbol{\beta_{hl}} \|_1 + (1 - a) \| \boldsymbol{\beta_{hl}} \|_2^2 \right],
$$
where $\lambda_{hl} > 0$ is the regularization parameter for transition $h \rightarrow l$, and $a \in [0, 1]$ controls the balance between $\ell_1$ (LASSO) and $\ell_2$ (ridge) penalization \citep{zou_regularization_2005}.
The proposed estimation algorithm combines an inner loop to estimate $\boldsymbol{\theta}$ and $\boldsymbol{\beta}$ for fixed values of the hyperparameters $(a,\boldsymbol{\lambda})$ and an outer loop to select the hyperparameters.
\subsubsection{Estimation procedure for known penalty parameters (Inner loop)}\label{innerloop}
For fixed penalty parameters $(a,\boldsymbol{\lambda})$, the inner loop maximizes the regularized log-likelihood over the regression parameters $\boldsymbol{\beta}$ and baseline transition intensity parameters $\boldsymbol{\theta}$. This inner loop is an iterative two-step algorithm combining at each iteration a cyclical gradient descent method to update the regression parameters $\beta$ and a Newton-Raphson-like algorithm to update the baseline intensity parameters $\boldsymbol{\theta}$. We note $(\boldsymbol{\widehat\theta},\boldsymbol{\widehat\beta}) = (\boldsymbol{\widehat\theta}(a,\boldsymbol{\lambda}),\boldsymbol{\widehat\beta}( a,\boldsymbol{\lambda}))$ the solution given by the last iteration of the inner loop. \\

{\noindent \it Update of regression parameters: } At iteration $(m+1)$, using the proximal coordinate gradient descent approach proposed by \cite{simon_regularization_2011}, each regression parameter $\beta_k$ is updated by maximizing the regularized log-likelihood over $\beta_k$ with $\boldsymbol{\theta}$ and $\boldsymbol{\beta_{-k}}$ (the vector $\boldsymbol{\beta}$ excluding the k-th element) fixed at their last values $\boldsymbol{\widehat \theta^{(m)}}$ and $\boldsymbol{\widehat\beta_{-k}^{(m)}}$:
\begin{equation} \label{eq:5}
\widehat\beta^{(m+1)}_k=\underset{\beta_k}{\argmax}\{\widetilde \ell( \beta_k |\boldsymbol{\widehat \theta^{(m)},\widehat\beta_{-k}^{(m)}})-\pen(\beta_k,\boldsymbol{\widehat\beta_{-k}^{(m)}},a,\boldsymbol{\lambda})\}
\end{equation} 
where $\widetilde \ell( \boldsymbol{\beta} |\boldsymbol{\widehat \theta^{(m)}},\boldsymbol{\widehat\beta_{-k}^{(m)}})$ is the second order Taylor development of the log-likelihood around $\boldsymbol{\widehat \beta^{(m)}}$ at fixed values $(\boldsymbol{\widehat \theta^{(m)}},\boldsymbol{\widehat\beta_{-k}^{(m)}})$ of the other parameters. For given 0$ \leq a \leq $1 and $\lambda_{hl}>$0, the solution of equation (\ref{eq:5}), when the $k^{th}$ element of $\boldsymbol{\beta}$ is associated to transition $h$ to $l$, is
\begin{align*}
    \widehat\beta_k^{(m+1)} \longleftarrow \frac{S(\displaystyle{ \nabla_k \ell(\boldsymbol{\widehat \beta^{(m)}},\boldsymbol{\widehat \theta^{(m)}}) - \widehat \beta_k^{(m)} \widehat x_{kk},a\lambda_{hl})}}{2\lambda_{hl}(1-a)-\widehat x_{kk}}
\end{align*}
where  $\nabla_k \ell$ the first derivative of the log-likelihood with respect to $\beta_k$ computed at $(\boldsymbol{\widehat \beta^{(m)}} , \boldsymbol{\widehat \theta^{(m)}})$, $\widehat x_{kk}$ the $k^{th}$ diagonal element of the Hessian matrix such that $\widehat x_{kk}=\nabla^2_{kk} \ell(\boldsymbol{\widehat \beta^{(m)}} , \boldsymbol{\widehat \theta^{(m)}})$ (note that if the Hessian matrix opposite is not definite positive, it is inflated according to Marquardt-Levenberg algorithm, \cite{philipps_robust_2021}) and the soft-threshold operator $S$ defined as $S(x,\lambda)=\sgn(x)(|x|-\lambda)_+$ which is equal to $x-\sgn(x)\lambda$ if $|x| > \lambda$ and 0 otherwise ($\sgn(x)=1$ if $x>0$; $\sgn(x)=-1$ if $x<$0 and 0 otherwise). Details on the analytical derivatives are provided in the Web Appendix. \\

\noindent {\it Update of baseline transition intensities parameters:} At iteration $(m+1)$, after having updated the $\boldsymbol{\beta}$ parameters, the update of parameters $\boldsymbol{\theta}$ is obtained by maximizing the log-likelihood over $\boldsymbol{\theta}$ with $\boldsymbol{\beta}$ fixed at its last value $\boldsymbol{\widehat \beta^{(m+1)}}$. We use a few iterations of a robust Newton-like algorithm, the Marquardt-Levenberg algorithm \citep{marquardt_algorithm_1963}, for which good convergence properties have been shown in various contexts of maximum likelihood estimation \citep{philipps_robust_2021}. \\

\noindent {\it  Stopping criteria for the inner loop:} The iterative procedure, which alternately updates $\boldsymbol{\beta}$ and $\boldsymbol{\theta}$, stops when parameter stability
\begin{equation*}
    \displaystyle{\lVert \boldsymbol{\widehat \beta^{(m+1)}} - \boldsymbol{\widehat\beta^{(m)}} \rVert_2^2 + \lVert \boldsymbol{\widehat \theta^{(m+1)}} - \boldsymbol{\widehat \theta^{(m)}}\rVert_2^2} < e_a
    \end{equation*}
and relative objective function stability  \begin{equation*}
    \displaystyle{\frac{| \: \ell_{pen}(\boldsymbol{\widehat \theta^{(m+1)}},\boldsymbol{\widehat \beta^{(m+1)}}) - \ell_{pen}(\boldsymbol{\widehat \theta^{(m)}},\boldsymbol{\widehat \beta^{(m)}}) \: |}{| \: \ell_{pen}(\boldsymbol{\widehat \theta^{(m)}},\boldsymbol{\widehat \beta^{(m)}}) \: |} < e_b}
\end{equation*}
 are both met. Here, $e_a$ and $e_b$ are user-defined thresholds that control the convergence tolerance ($10^{-5}$ by default for both).
 
\subsubsection{Selection of penalty parameters (Outer loop)}\label{pena_choice}
After the inner loop convergence, the pair $(\boldsymbol{\widehat \theta},\boldsymbol{\widehat \beta})=(\boldsymbol{\widehat \theta}(a,\boldsymbol{\lambda}),\boldsymbol{\widehat \beta}(a,\boldsymbol{\lambda}))$ represents the estimated baseline intensity parameters and regression coefficients for fixed penalty parameters $(a,\boldsymbol{\lambda})$. We denote by $\boldsymbol{\mathcal I(\widehat \beta( a,  \lambda))}$ the active set of covariates - that is, those associated with non-zero regression coefficients - and by $\big|\boldsymbol{\mathcal I(\widehat \beta( a,  \lambda))}\big|$ its cardinality. The four penalty parameters are then selected as those that minimize the Bayesian Information Criterion (BIC) \citep{neath_bayesian_2012} through a grid search. The BIC is defined as
\begin{align*}
 (\boldsymbol{\widehat \theta}(\hat a, \boldsymbol{\hat \lambda}), \boldsymbol{\widehat \beta}(\hat a,\boldsymbol{\hat \lambda})) & = \underset{\boldsymbol{\theta}(a,\boldsymbol{\lambda}),\boldsymbol{\beta}(a,\boldsymbol{\lambda})}{\argmin}\{BIC(\boldsymbol{\theta}(a,\boldsymbol{\lambda}),\boldsymbol{\beta}(a,\boldsymbol{\lambda}))\} \\
 & = \underset{a,\boldsymbol{\lambda}}{\argmin}\{-2\ell(\boldsymbol{\widehat \theta}(a,\boldsymbol{\lambda}), \boldsymbol{\widehat \beta}(a,\boldsymbol{\lambda}))+\log(N)\big|\boldsymbol{\mathcal I(\widehat \beta( a,  \lambda))}\big|\}.
\end{align*}

To reduce the computational cost of the grid search for four penalty parameters, we use a structured pre-selection strategy. We first optimize each transition-specific penalty parameter $(\lambda_{01}, \lambda_{02}, \lambda_{12})$ independently using regularized models with covariates only on the transition of interest. In the rest of the manuscript, for each of the four values of $a \in \{0.25, 0.50, 0.75, 1\}$, we evaluate 20 values of $\lambda_{hl}$ for each transition $hl$, retaining the three best $\lambda_{hl}$ values based on BIC. This results in 108 candidate combinations of $(a, \lambda_{01}, \lambda_{02}, \lambda_{12})$, from which the one minimizing the BIC for the full model including covariates on the three transitions is finally selected. The R code of the complete estimation algorithm, including penalty parameters selection is implemented in the R package HIDeM \citep{HIDeM}. 
Following, we denote $\boldsymbol{\mathcal I(\widehat \beta(\hat a, \hat \lambda))}$ the final set of selected covariates. 

\subsection{Post-selection estimation}\label{coveffect}
After performing variable selection,  we refit an unpenalized IDM with $\boldsymbol{\mathcal I(\widehat \beta(\hat a, \hat \lambda))}$ and estimate the corresponding regression parameters $\boldsymbol{\widehat \beta^{\MLE}}$ and baseline intensity parameters $\boldsymbol{\widehat \theta^{\MLE}}$ via maximum likelihood estimation \citep{belloni_least_2013}. The log-likelihood is maximized using the Marquardt-Levenberg algorithm, and the covariance matrix of parameter estimates is estimated by the inverse of the negative Hessian matrix. 

\newpage
\section{Simulations}\label{simu} 
We followed the recommendations of \cite{morris_using_2019} to report the simulation study.

\subsection{Aim} 
The simulation aimed to evaluate the performance of the regularized IDM in terms of variable selection and prediction accuracy considering 6 scenarios varying according to the interval between visits, the rates of events and the correlation structure between covariates. We also compared its performance with that of a PHM for semi-competing risks, which neglects interval-censoring of the disease onset time.

\subsubsection{Data generation mechanism}
We simulated follow-up data for samples of 
2,000 individuals. All individuals entered the study in the healthy state ($t$=0), and follow-up ended administratively at $t$=18 years. Between these two time points, follow-up visits were scheduled on a fixed time grid, with two follow-up frequencies: either one visit every 2.5 years (up to 8 visits) or every 4.5 years (up to 5 visits) leading to frequent or sparse visit schedule, respectively. To mimic the irregularity typically observed in cohort studies, actual visit times were jittered by adding random uniform noise between 0 and +0.5 years around the scheduled times. Non-informative censoring was also added with a 5$\%$ dropout probability at each visit. 

For each individual $i$, we generated a set of covariates $\boldsymbol{Z_i}$ comprising 50 standardized zero-mean Gaussian variables, with two covariance structures: independent or correlated with a group-Toeplitz correlation structure of 0.5 (see heatmap in Figure~\ref{heatmap} Web Appendix). The covariates were grouped into five blocks of ten variables, with only three of them associated with at least one of the transition intensities (see Table~\ref{covariate_effect} in Web Appendix). 

The individual transition times $T^{hl}_i$ were generated using the inversion method assuming transition-specific Weibull distributions. To assess the impact of the event rates, we considered low and high event rates—approximately 15\% and 35\% for illness, and 35\% and 80\% for death, respectively—by specifying different values for the Weibull baseline intensity parameters (see Table~\ref{scenarios_descript} in the Web Appendix). We then derived the true time to disease onset $T^I_i = \min\{T^{01}_i, T^{02}_i, 18\}$, and the corresponding true disease indicator $\delta_{i}^{I*}$ ($\delta_{i}^{I*}=1$ if $T^I_i = T^{01}_i$, 0 otherwise). Since the time of disease onset was interval-censored, we defined $L_i$ as the time of the last visit at which the individual was observed healthy (i.e. the last visit before $T^I_i$), and $R_i$ as the time of the visit at which the diagnosis was made —that is, the first visit after $T^I_i$. The observed disease indicator $\delta_{i}^{I}$ was then given by
$$
\delta_{i}^{I} = 
\begin{cases}
1 & \text{if } \delta_{i}^{I*} = 1,\, R_i \geq T^I_i,\, \text{and } R_i < \min\{T^{12}_i, 18\}, \\
0 & \text{otherwise}.
\end{cases}
$$
As death times were fully observed up to the administrative end of follow-up, the individual time to death was defined as:
$$
T^D_i = 
\begin{cases}
\min\{T^{12}_i, 18\} & \text{if } T^I_i = T^{01}_i, \\
\min\{T^{02}_i, 18\} & \text{otherwise},
\end{cases}
$$
with the corresponding death indicator
$$
\delta_{i}^{D} = 
\begin{cases}
\mathds{1}_{\{T^{12}_i < 18\}} & \text{if } \delta^{I*}_i = 1, \\
\mathds{1}_{\{T^{02}_i < 18\}} & \text{otherwise}.
\end{cases}
$$

This data generation setting is summarized in Table~\ref{scenarios_descript} Web Appendix, we investigated six scenarios for data generation denoted A.1, A.2, B.1, B.2, C.1 and C.2: low event rates and frequent visit (A), high event rates and frequent visit (B) and high event rates and sparse visit (C) with independent (.1) or group-Toeplitz (.2) correlation structure between covariates.

\subsubsection{Estimands} \label{estimands}

Each data set was composed of a training sample of 2,000 subjects and a test sample of 2,000 subjects. The estimands of interest were the active set of covariates $\boldsymbol{\mathcal I(\widehat \beta(\hat a, \hat \lambda))}$ on the training sample and the individual predicted probability to develop dementia before the time of first event ($t_i=\min\{T^{01}_i,T^{02}_i,18\}$) for subjects in the test sample, noted $\widehat F_{01i}(t_i)$:
\begin{equation*}
    \hat F_{01i}(t_i)=\int_{0}^{t_i}e^{-\hat A_{01i}(u)-\hat A_{02i}(u)}\hat \alpha_{01i}(u)du
\end{equation*}
with $e^{-\hat A_{01i}(u)-\hat A_{02i}(u)}$ the estimated probability of staying alive and healthy until time $u$ and $\hat \alpha_{01i}(u)$ the estimated transition intensity for transition from {\em healthy} to {\em diseased} at time $u$ using the refitted estimation $(\boldsymbol{\widehat\theta^{\MLE}},\boldsymbol{\widehat\beta^{\MLE}})$ - see Section~\ref{coveffect}. Note that the true value 
of $F_{01i}(t_i)$ can be obtained using the true parameter values. 
\newpage
\subsubsection{Methods}\label{models}
For each generated training sample, we estimated the following models: 
\begin{itemize}
     \item Regularized IDM-ICT: our proposed method, an IDM handling Interval-Censored Time to disease onset (ICT) with variable selection. 
    \item Oracle IDM-ICT: an (unregularized) IDM handling ICT with only the relevant covariates on each transition. It was the benchmark model as it mimicked a perfect selection of variables and handled interval-censored data.
    \item Regularized IDM-TT: a regularized IDM considering the exact True Time to disease onset and performing variable selection. This model could identify the active set of covariates without suffering from the complexity of interval-censoring; it reflected the variable selection benchmark for regularized IDM-ICT.  
    \item Regularized PHM: a semi-competing risk PHM performing variable selection on {\em healthy-diseased} and {\em healthy-deceased} transitions with the time of event defined as the mid-point between the last visit illness-free and the diagnosis visit for ill individuals. Following \cite{Proust2019}, subjects who died less than 3 years after the last visit disease-free were considered as dead disease-free while those who died later were considered as right censored at their last visit disease-free. 
\end{itemize}
All models were estimated with HIDeM assuming an M-spline basis for the baseline transition intensities (knots at 0, 9, and 18 years) and stopping criteria $e_a=10^{-7}$ and $e_b=10^{-4}$.

\subsubsection{Performance assessments}

For each scenario, the procedure was repeated across 500 simulated datasets. Performance was assessed in terms of variable selection accuracy and predictive ability. For each transition, we calculated the proportion of replicates in which each variable was selected and evaluated this frequency against the known relevance of the variable. The predictive ability was evaluated on the test set by the Mean Square Error of the Probability (MSEP) of becoming diseased before the individual censored time of the first-event $ t_i$=min\{$T^{01}_i,T^{02}_i,18$\}:
\begin{equation*}
    \widehat{MSEP}=\frac{1}{N}\sum_{i=1}^{N}(\widehat F_{01i}(t_i) - F_{01i}(t_i))^2.
\end{equation*}

\newpage
\subsubsection{Results}

\begin{center}
\begin{figure}[H]
    \centering
   \hspace*{-1.5cm}\includegraphics[width=18cm,height=14cm]{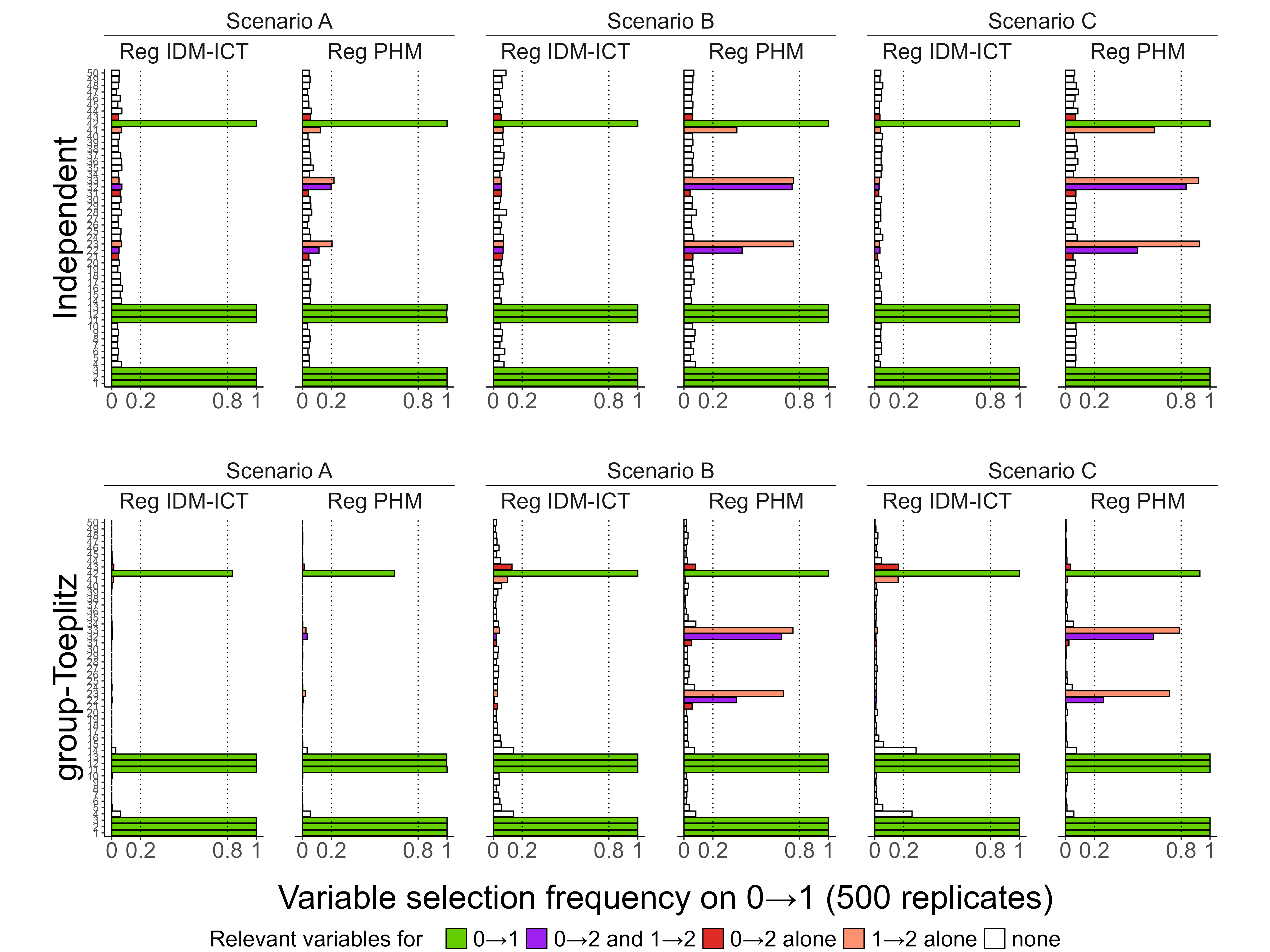}
    \caption{Variable selection frequency over 500 simulated replicates for the transition  {\em healthy} to {\em illness}, using Reg IDM-ICT and Reg PHM with M-spline basis for baseline transition, under low event rates with a frequent visit schedule (A) or high event rates with a frequent or sparse visit schedule (B and C) and two correlation structures among the covariates (Independent or group-Toeplitz).}
    \label{selection_01}
\end{figure}
\end{center}

Reg IDM-ICT demonstrated great variable selection performance across all scenarios (Figure \ref{selection_01}). Covariates truly associated with the disease were consistently selected across all 500 training samples, while those associated only with death or irrelevant for the 3 transitions were rarely selected (selection frequency $<$ 5\% in most cases). In contrast, under high event rates (Scenario B and C) the Reg PHM incorrectly selected covariates to be associated with death after illness, with frequencies ranging from 35\% to 75\% in Scenario B and reaching up to 90\% in Scenario C.

For the transition from {\em healthy} to  {\em deceased} (Figure~\ref{selection_02} in Web Supplementary), across all scenarios both Reg IDM-ICT and Reg PHM succeeded in identifying the covariates associated with illness-free death, but under high event rates (Scenarios B and C) the Reg PHM incorrectly selected covariates associated with death after illness only, in at least 50\% of the replicates. 
The variable selection performances of the proposed Reg IDM-ICT were very similar to those of the Reg IDM-TT, in which the true event times were known (Figures ~\ref{selection_01_true} and ~\ref{selection_02_true} in Web Supplementary). True positive rate and false positive rate of each method are further reported in Web Supplementary  Table~\ref{TPR-FPR}. \\

\begin{figure}[H]
    \centering
    \hspace*{-12ex}\includegraphics[width=18cm]{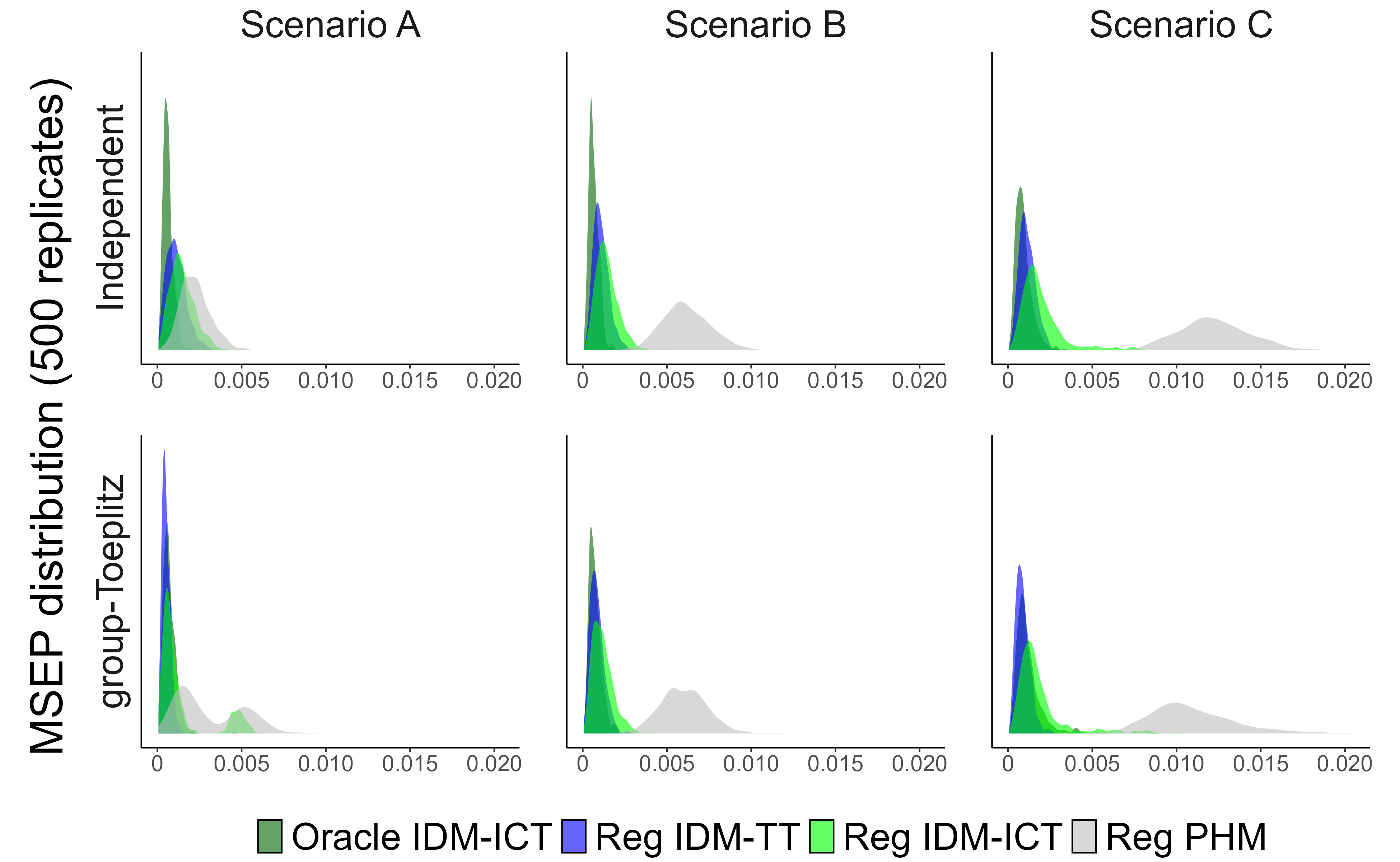}
    \caption{Distribution over 500 replicates of the Mean Square Error of the Probability (MSEP) for a healthy subject to become ill, using the Oracle IDM-ICT ({\color{green(ryb)}dark green}), the Reg IDM-TT ({\color{blue}blue}), Reg IDM-ICT ({\color{green(light)}green}) and the Reg PHM ({\color{lightgray}grey}) with M-spline basis for baseline transition, under low event rates with a frequent visit schedule (A) or high event rates with a frequent or sparse visit schedule (B and C) and two correlation structures among the covariates (Independent or group-Toeplitz).}
    \label{MSEP}
\end{figure}

The MSEP distribution obtained with our method -Reg IDM-ICT- closely matched those of Oracle IDM-ICT (model including only relevant variables) and Reg IDM-TT (variable selection using true disease times) (Figure \ref{MSEP}). In contrast, under high event rates (Scenario B and C) Reg PHM, which neglects the interval censoring, showed markedly larger MSEP values. Under high event rates and frequent visit schedule (Scenario B), the MSEP of the Oracle IDM-ICT ranged from 1x$10^{-4}$ to 3x$10^{-3}$, with a mean value of 7x$10^{-4}$ on independent and group-Toeplitz covariates. 

While our proposed Reg IDM-ICT followed closely by achieving a MSEP distribution between 2x$10^{-4}$ and 5x$10^{-3}$ with a mean of 1x$10^{-3}$ for both correlation scenarios, the Reg PHM had poorer performance with values ranging from 2x$10^{-3}$ to 1x$10^{-2}$ and a mean at 6x$10^{-3}$ for both correlation scenarios. Under high event rates and sparse visit schedule (Scenario C), the gap between the two methods became even more pronounced. Reg IDM-ICT maintained stable performance in regards to Oracle IDM-ICT, while the Reg PHM performance worsened with a mean value at 1x$10^{-2}$ for both independent and group-Toeplitz covariates.  \\

Overall, these results demonstrate that under high event rates (Scenario B and C) Reg IDM-ICT outperformes the Reg PHM in terms of both variable selection and predictive accuracy. However, under low event rates and frequent visit (Scenario A), the performances of both approaches tend to be similar. More importantly, the Reg IDM-ICT provides reliable variable selection and accurate predictions even in challenging settings characterized by high interval censoring and strong correlation between potential predictors. Similar findings for variable selection and predictive performance were observed when assuming Weibull baseline transition intensities instead of M-splines (Figures ~\ref{selection_weib_01},  ~\ref{selection_weib_02} and ~\ref{MSEP_weib} in Web supplementary).

\newpage
\vspace*{\fill}
\vspace*{\fill}

\newpage

\section{Application} \label{appli}

\subsection{ Population-based data from the Three-City (3C) study}
The Three-City (3C) study is a French prospective population-based study designed to investigate the association between vascular diseases and dementia in the elderly \citep{the_3c_study_group_vascular_2003}. Individuals aged 65 years and older were randomly enrolled from the electoral lists of three French cities (Bordeaux, Dijon and Montpellier) in 1999. A total of 9,294 subjects underwent extensive health examinations and baseline risk factor assessment, and were subsequently followed through visits planned every 2 or 3 years up to 20 years. The present study was restricted to subjects from the Bordeaux and Dijon centers who underwent MRI examination at baseline. Among the 7,035 subjects enrolled in these two centers, 2,199 had a measure of MRI markers, and after exclusion of 220 subjects with missing information on at least one medical or neuropsychological evaluation, the final sample comprised 1,979 subjects. \\

For the 1,979 subjects, we considered a set of 27 potential predictors collected at baseline, including 3 socio-demographic characteristics, 10 health-related markers, 5 cognitive test scores, one genetic factor (presence of at least one $\epsilon$4 allele of the Apolipoprotein E genotype ApoE4), 5 brain volume measures and 4 markers of cerebral Small Vessel Disease markers (cSVD), both derived from the brain MRI. \\

Socio-demographic characteristics included age at study entry, sex (female vs male) and education level, which was classified as short (secondary school or less) or long (more than secondary school). 

Health-related information collected at baseline included two binary indicators of functional impairment: the Rosow and Breslow mobility scale (MOBILITY, defined as impaired if the mobility was restricted in walking between 500m and 1km or stair climbing) and the Instrumental Activity of Daily Living (IADL, considered impaired if the subject was not independent in using the telephone, transportation, medication management, or budgeting). 

Clinical measures included Systolic and Diastolic Blood Pressure (SBP and DBP, mean of two evaluations in a seated position), Body Mass Index (BMI), depressive symptomatology score using the Center for Epidemiologic Studies-Depression scale (CESD), medication intake (\#Drugs intake in the last month) and comorbidities history of diabetes (under treatment or blood glucose level $\geq$ 7 mmol/L), coronary disease (myocardial infarction or angina pectoris or coronary surgery) and High Blood Pressure (HBP, under treatment or SBP $\geq$ 140 mmHg or DBP $\geq$ 90 mmHg). 

Cognitive functioning was assessed by 5 cognitive tests: the Benton Visual Retention Test (Benton), the Mini-Mental State Examination (MMSE), the Trail Making Test part A and B (TMT A and B) and the Isaacs’s Set Test (IST). 

We considered 5 major brain volumetric variables (Total Intracranial Volume -TIV, Grey Matter Volume -GMV, White Matter Volume -WMV, hippocampus volume and medial temporal lobe volume) as well as 4 cSVD markers (deep and perivascular White Matter Hyperintensities -WMH, Deep White Matter -DWM and basal ganglia PeriVascular Spaces -PVS). 

\subsection{Diagnosis of dementia and death}
Dementia diagnosis was evaluated at each visit according to the Diagnostic and Statistical Manual of Mental Disorders, Fourth Edition revised \citep{vahia_diagnostic_2013}, and using a three-step procedure. A psychologist screened subjects based on their neuropsychological performance and their change relative to previous evaluations. Subjects suspected of having dementia were redirected to a senior neurologist who established the clinical diagnosis. Final diagnosis was established by a committee of experts reaching a consensus on the diagnosis. Alive status and exact date of death were collected continuously throughout the follow-up.

\subsection{Description of the sample}
Among the 1,979 subjects included in the analysis, 1,216 (61\%)  were  women,  813 (41\%) went to secondary school or higher, and 430 (22\%) had at least one ApoE $\epsilon$4 allele (Table~\ref{descrip3C} in Web Supplementary). Only a few subjects showed dependency at baseline (4\% and 6\%, respectively, for mobility and IADL), 8\% had a history of diabetes, 5\% of coronary artery disease, and 38\% of hypertension. Subjects entered the study at 72.3 years old on average (SD=4). They were followed up for 9.1 years on average with a number of visits per subject ranging from 1 to 8. Along the follow-up, 198 (10\%) subjects were diagnosed with dementia, and 89 (4\%) of them died; 403 (20\%) subjects died without dementia diagnosis, and 1,378 (70\%) were alive dementia-free at the end of the study.

\subsection{Results}
\begin{figure}[H]
    \hspace{-12ex}\includegraphics[width=18cm,height=12cm]{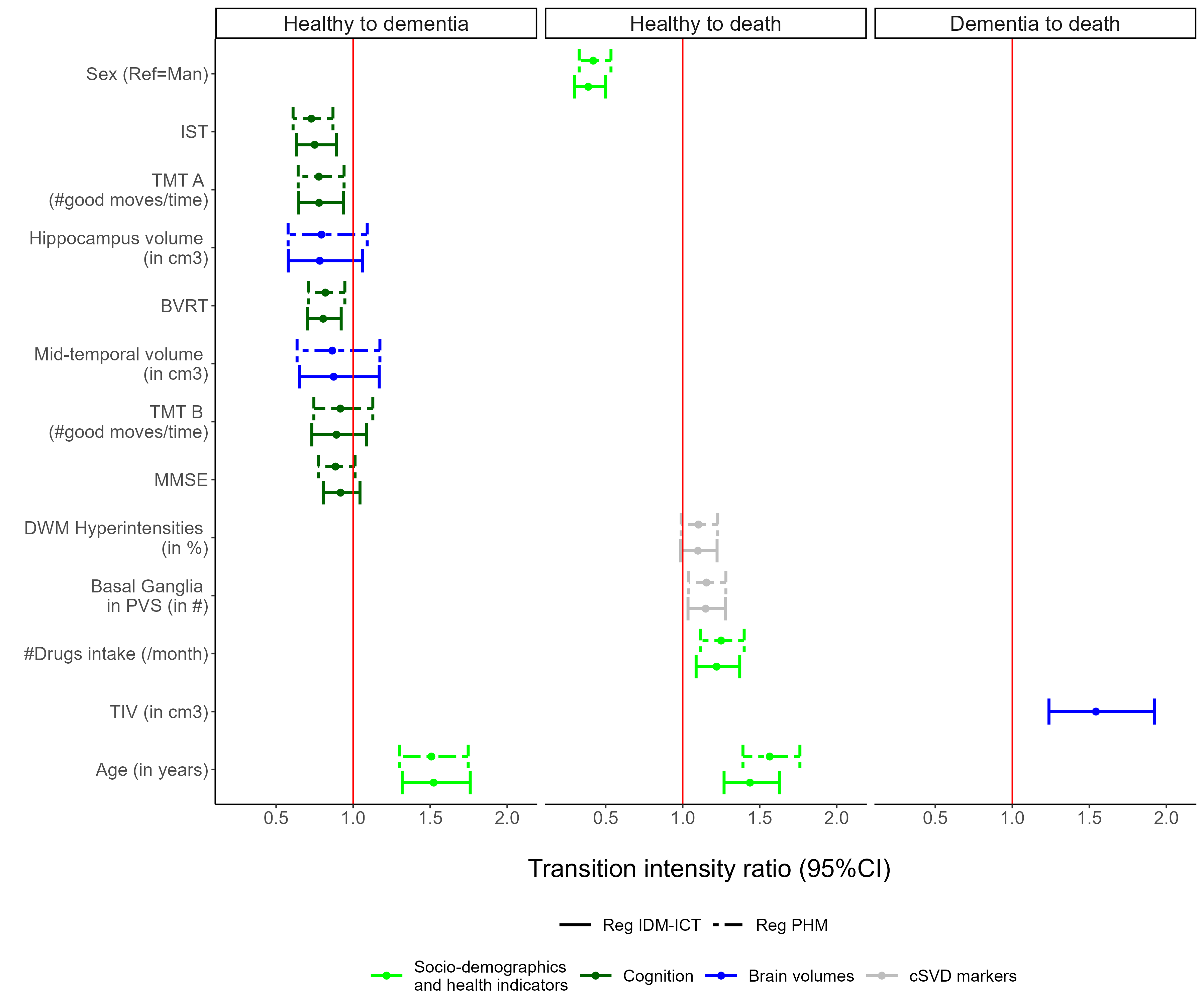}
    \caption{Estimated transition intensity ratios (and 95\% confidence interval) of each covariate for transitions from healthy to dementia, from healthy to dead and from dementia to dead in the 3C Study (N=1,979).}
    \label{HR3C}
     \caption*{\small \textit{Abbreviations:}  MMSE=Mini-Mental State Examination, TMT=Trail Making Test, TIV=Total Intracranial Volume, DWM= Deep White matter and PVS=PeriVascular Spaces.}
\end{figure}

Figure \ref{HR3C} displays the estimated transition intensity ratios from the refitted IDM-ICT and PHM with M-spline basis for the baseline transition (knots at 0, 5, 9 and 18 years), including only the predictors selected by the regularized approach. Both our proposed Reg IDM-ICT and the semi-competing Reg PHM selected the same set of predictors and yielded similar estimates of transition intensity ratios. 

For the transition from {\em healthy state} to {\em dementia}, age at entry, the 5 cognitive tests, as well as hippocampus and medial temporal lobe volumes, were identified as predictors. As expected,  older ages, lower processing speed (TMT A), reduced verbal fluency (IST), and poorer visual memory (Benton) were significantly associated with higher transition intensity to dementia. 

For the transition from {\em healthy} to {\em deceased}, male sex, older ages, higher number of medications and higher counts of PVS in the basal ganglia were significantly associated with an increased transition intensity to death dementia-free. 

For {\em death} after {\em dementia}, unexpectedly only TIV was selected as a predictor. This may suggest that subjects with greater brain reserve and having clinical signs of dementia tended to experience a more rapid deterioration in overall health status \citep{foubert-samier_education_2012}.

\begin{sidewaysfigure}
\includegraphics[height=16cm]{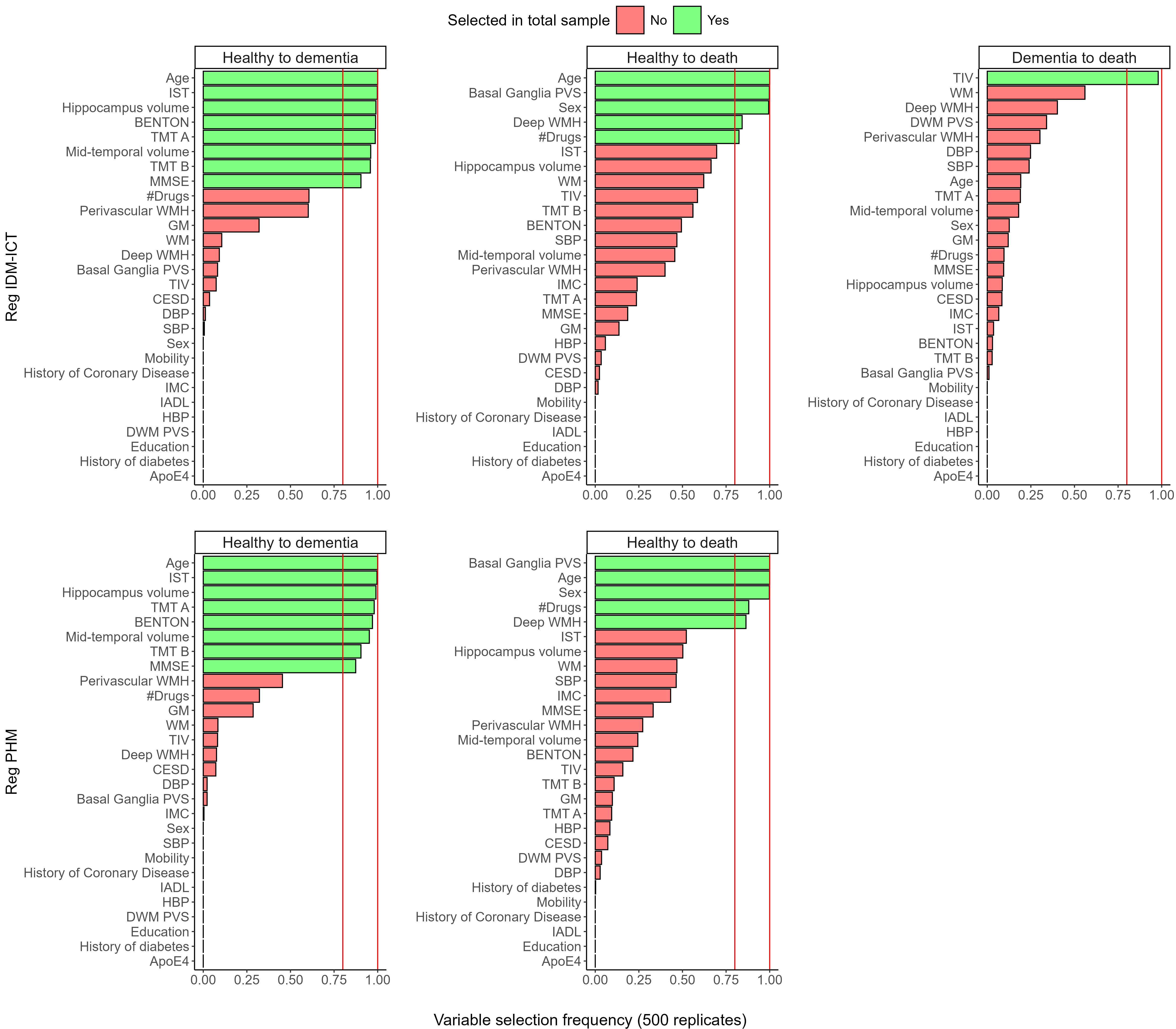}
     \caption{Proportion of selection over 500 bootstrap samples of each covariate on the three transitions using Reg IDM-ICT and Reg PHM with M-spline basis for baseline transition in 3C Stduy (N=1,979)}
    
    \caption*{\small \textit{Abbreviations:} HBP = High Blood Pressure, BMI = Body Mass Index, CESD = Center for Epidemiologic Studies Depression Scale, IADL = Instrumental Activities of Daily Living, MMSE = Mini-Mental State Examination, TMT = Trail Making Test, TIV = Total Intracranial Volume, WM = White Matter, GM = Grey Matter, WMH = White Matter Hyperintensities, PVS = Perivascular Spaces.}
    \label{boot_3C}
\end{sidewaysfigure}

We replicated the variable selection over 500 bootstrap samples  \citep{altman_bootstrap_1989} (with penalty parameters fixed at the values selected on the total sample). For both methods, this highlighted the very good stability of the variable selection for transition from {\em healthy} to {\em dementia} and from {\em dementia} to {\em death} and a quite good stability for transition from {\em healthy} to {\em death} (Figure \ref{boot_3C}).

\section{Discussion}  \label{discussion}

Regularized survival analysis has previously been developed to perform variable selection and account for an uncertain time to disease onset \citep{li_adaptive_2020,zhao_simultaneous_2020,feng_variable_2023}. However, to our knowledge, none of these regularized approaches have accounted for the undiagnosed disease cases due to the combination of interval-censoring and semi-competing risk of death. To tackle this challenge, we have propose an estimation procedure based on penalized likelihood for an illness–death model with proportional transition intensities accounting for interval censoring.

We proposed an efficient optimization approach using both proximal Gradient descent and the Marquardt-Levenberg algorithm. By iteratively using the two algorithms, we accurately estimated the regression parameters and baseline transition intensity parameters for each transition without ignoring the correlation between them. In addition, the methodology has been implemented in an R package HIDeM —with detailed replicable code— to provide a user-friendly function enabling scientists to carry out variable selection in high dimensional settings on interval-censored data \citep{HIDeM}. Although this work focused on elastic-net penalty, HIDeM package also implements Minimax Concave Penalty and Smoothly Clipped Absolute Deviation penalty \citep{MCP_SCAD} following the same strategy.

 When the rate of events is high or the interval between visits is large, the simulation study highlighted the superiority of the proposed approach over the classical regularized semi-competing risk approach which neglects interval censoring. When interval censoring is neglected, the regularized semi-competing risk model improperly selects in the disease transition submodel predictors that are only associated with death risk among diseased subjects. The proposed approach avoids these errors and consequently leads to much better predictive ability, close to the oracle model in the simulation study. In the application, our approach successfully identified known predictors of dementia separately from predictors of death with or without a dementia diagnosis. However in this data set, events rates and visit schedule are comparable to simulation Scenario A, where the impact of interval censoring was low, and as expected, the illness-death model and the classical semi-competing risk model yielded the same variable selection.

The proposed regularized illness-death model presents three main limitation. First, the methodology has been proposed for time-fixed covariates. Second, using proportional transition intensities permitted to have an easy interpretation of the covariates parameters but implied a strong proportionality assumption over 20 years of follow-up. Finally, to perform post-selection inference, we fitted an IDM-ICT with only the active set of covariates. Thus, by omitting the selection variability, we likely underestimated the variance of the regression parameters \citep{rasines_splitting_2023}. To the best of our knowledge, implementing unbiased variance estimates in this context is not straightforward, we leave this implementation to future methodological development.

\section*{Acknowledgments}
The authors thank the 3C study investigators for access to the Bordeaux and Dijon data. Computer time for this study was provided by the computing facilities MCIA (M\'esocentre de Calcul Intensif Aquitain) of the Universit\'e de Bordeaux and of the Universit\'e de Pau et des Pays de l’Adour. This work was supported by a French government grant managed by the Agence Nationale de la Recherche under the France 2030 program, reference ANR-22-PESN-0016 REWIND and carried out within the University of Bordeaux's France 2030 program/RRI PHDS. This project was also partly supported by a grant overseen by the French National Research Agency (ANR) as part of the France 2030 plan ANR-18-RHUS-0002.

\section*{Supplementary materials}
HIDeM R package is available on GitHub \citep{HIDeM}. Web Appendices, Tables, and Figures referenced in Sections \ref{methodo}, \ref{simu} and \ref{appli} are available with this paper.
\section*{Data availability}
Anonymized 3C data may be shared upon reasonably justified request to the 3C scientific committee (e3C.CoordinatingCenter@u-bordeaux.fr).

\bibliography{biblio}

\begin{thebibliography}{10}

\bibitem{altman_bootstrap_1989}
D.~G. Altman and P.~K. Andersen.
\newblock Bootstrap investigation of the stability of a cox regression model.
\newblock {\em Statistics in Medicine}, 8(7):771--783, July 1989.

\bibitem{andersen_competing_2002}
P.~K. Andersen, S.~Z. Abildstrom, and S.~Rosthøj.
\newblock Competing risks as a multi-state model.
\newblock {\em Statistical Methods in Medical Research}, 11(2):203--215, Apr.
  2002.

\bibitem{andersen_competing_2012}
P.~K. Andersen, R.~B. Geskus, T.~De~Witte, and H.~Putter.
\newblock Competing risks in epidemiology: possibilities and pitfalls.
\newblock {\em International Journal of Epidemiology}, 41(3):861--870, June
  2012.

\bibitem{belloni_least_2013}
A.~Belloni and V.~Chernozhukov.
\newblock Least squares after model selection in high-dimensional sparse
  models.
\newblock {\em Bernoulli}, 19(2), May 2013.

\bibitem{HIDeM}
A.~Bercu, A.~Guilloux, H.~Jacqmin-Gadda, and C.~Proust-Lima.
\newblock Hidem: In high dimension, estimation of smooth hazard models for
  interval-censored data with applications to survival and illness-death
  models.
\newblock \url{https://github.com/arianebercu/HIDeM}, 2025.
\newblock Accessed: 2025-06-06.

\bibitem{MCP_SCAD}
P.~Breheny and J.~Huang.
\newblock {Coordinate descent algorithms for nonconvex penalized regression,
  with applications to biological feature selection}.
\newblock {\em The Annals of Applied Statistics}, 5(1):232 -- 253, 2011.

\bibitem{chen_interval-censored_2012}
D.-G.~D. Chen, J.~Sun, and K.~E.~E. Peace.
\newblock {\em Interval-{Censored} {Time}-to-{Event} {Data}: {Methods} and
  {Applications}}.
\newblock Chapman and Hall/CRC, New York, July 2012.

\bibitem{commenges_dynamical_2015}
D.~Commenges and H.~Jacqmin-Gadda.
\newblock {\em Dynamical {Biostatistical} {Models}}.
\newblock Chapman and Hall/CRC, New York, Oct. 2015.

\bibitem{feng_variable_2023}
F.~Feng, G.~Cheng, and J.~Sun.
\newblock Variable {Selection} for {Length}-{Biased} and {Interval}-{Censored}
  {Failure} {Time} {Data}.
\newblock {\em Mathematics}, 11(22):4576, Nov. 2023.

\bibitem{fine_semi-competing_2001}
J.~P. Fine, H.~Jiang, and R.~Chappell.
\newblock On semi-competing risks data.
\newblock {\em Biometrika}, 88(4):907--919, Dec. 2001.

\bibitem{foubert-samier_education_2012}
A.~Foubert-Samier, G.~Catheline, H.~Amieva, B.~Dilharreguy, C.~Helmer,
  M.~Allard, and J.-F. Dartigues.
\newblock Education, occupation, leisure activities, and brain reserve: a
  population-based study.
\newblock {\em Neurobiology of Aging}, 33(2):423.e15--423.e25, Feb. 2012.

\bibitem{the_3c_study_group_vascular_2003}
T.~C.~S. Group.
\newblock Vascular {Factors} and {Risk} of {Dementia}: {Design} of the
  {Three}-{City} {Study} and {Baseline} {Characteristics} of the {Study}
  {Population}.
\newblock {\em Neuroepidemiology}, 22(6):316--325, 2003.

\bibitem{joly_penalized_2002}
P.~Joly, D.~Commenges, C.~Helmer, and L.~Letenneur.
\newblock A penalized likelihood approach for an illness-death model with
  interval-censored data: application to age-specific incidence of dementia.
\newblock {\em Biostatistics}, 3(3):433--443, Sept. 2002.

\bibitem{leffondre_interval-censored_2013}
K.~Leffondré, C.~Touraine, C.~Helmer, and P.~Joly.
\newblock Interval-censored time-to-event and competing risk with death: is the
  illness-death model more accurate than the {Cox} model?
\newblock {\em International Journal of Epidemiology}, 42(4):1177--1186, Aug.
  2013.

\bibitem{li_adaptive_2020}
C.~Li, D.~Pak, and D.~Todem.
\newblock Adaptive lasso for the {Cox} regression with interval censored and
  possibly left truncated data.
\newblock {\em Statistical Methods in Medical Research}, 29(4):1243--1255, Apr.
  2020.

\bibitem{liang_mortality_2021}
C.-S. Liang, D.-J. Li, F.-C. Yang, P.-T. Tseng, A.~F. Carvalho, B.~Stubbs,
  T.~Thompson, C.~Mueller, J.~I. Shin, J.~Radua, R.~Stewart, T.~K. Rajji, Y.-K.
  Tu, T.-Y. Chen, T.-C. Yeh, C.-K. Tsai, C.-L. Yu, C.-C. Pan, and C.-S. Chu.
\newblock Mortality rates in {Alzheimer}'s disease and non-{Alzheimer}'s
  dementias: a systematic review and meta-analysis.
\newblock {\em The Lancet Healthy Longevity}, 2(8):e479--e488, Aug. 2021.
\newblock Publisher: Elsevier.

\bibitem{marquardt_algorithm_1963}
D.~W. Marquardt.
\newblock An {Algorithm} for {Least}-{Squares} {Estimation} of {Nonlinear}
  {Parameters}.
\newblock {\em Journal of the Society for Industrial and Applied Mathematics},
  11(2):431--441, June 1963.

\bibitem{morris_using_2019}
T.~P. Morris, I.~R. White, and M.~J. Crowther.
\newblock Using simulation studies to evaluate statistical methods.
\newblock {\em Statistics in Medicine}, 38(11):2074--2102, May 2019.

\bibitem{neath_bayesian_2012}
A.~A. Neath and J.~E. Cavanaugh.
\newblock The {Bayesian} information criterion: background, derivation, and
  applications.
\newblock {\em WIREs Computational Statistics}, 4(2):199--203, Mar. 2012.

\bibitem{philipps_robust_2021}
V.~Philipps, B.~Hejblum, P., M.~Prague, D.~Commenges, and C.~Proust-Lima.
\newblock Robust and {Efficient} {Optimization} {Using} a
  {Marquardt}-{Levenberg} {Algorithm} with {R} {Package} {marqLevAlg}.
\newblock {\em The R Journal}, 13(2):273, 2021.

\bibitem{Proust2019}
C.~Proust-Lima, V.~Philipps, and J.-F. Dartigues.
\newblock A joint model for multiple dynamic processes and clinical endpoints:
  Application to alzheimer's disease.
\newblock {\em Statistics in Medicine}, 38(23):4702--4717, 2019.

\bibitem{rasines_splitting_2023}
D.~G. Rasines and G.~A. Young.
\newblock Splitting strategies for post-selection inference.
\newblock {\em Biometrika}, 110(3):597--614, Aug. 2023.

\bibitem{simon_regularization_2011}
N.~Simon, J.~Friedman, T.~Hastie, and R.~Tibshirani.
\newblock Regularization {Paths} for {Cox}'s {Proportional} {Hazards} {Model}
  via {Coordinate} {Descent}.
\newblock {\em Journal of Statistical Software}, 39(5):1--13, 2011.

\bibitem{tibshirani_regression_1996}
R.~Tibshirani.
\newblock Regression {Shrinkage} and {Selection} {Via} the {Lasso}.
\newblock {\em Journal of the Royal Statistical Society Series B: Statistical
  Methodology}, 58(1):267--288, Jan. 1996.

\bibitem{vahia_diagnostic_2013}
V.~Vahia.
\newblock Diagnostic and statistical manual of mental disorders 5: {A} quick
  glance.
\newblock {\em Indian Journal of Psychiatry}, 55(3):220, 2013.

\bibitem{zhao_simultaneous_2020}
H.~Zhao, Q.~Wu, G.~Li, and J.~Sun.
\newblock Simultaneous {Estimation} and {Variable} {Selection} for
  {Interval}-{Censored} {Data} {With} {Broken} {Adaptive} {Ridge} {Regression}.
\newblock {\em Journal of the American Statistical Association},
  115(529):204--216, Jan. 2020.

\bibitem{zou_regularization_2005}
H.~Zou and T.~Hastie.
\newblock Regularization and {Variable} {Selection} {Via} the {Elastic} {Net}.
\newblock {\em Journal of the Royal Statistical Society Series B: Statistical
  Methodology}, 67(2):301--320, Apr. 2005.

\end{thebibliography}
\bibliographystyle{abbrv}

\newpage
\section*{Web Supplementary}
\appendix

\renewcommand{\thetable}{S\arabic{table}}
\renewcommand{\thefigure}{S\arabic{figure}}
\setcounter{figure}{0}
\setcounter{table}{0}

\section*{Analytical derivatives}
For individual i, who entered the study at $V_{0i}$, remained healthy until visit $L_i$, was diagnosed with the disease at visit time $R_i$ and was either censored at time $T_i$ (case 1) or died at $T_i$ (case 2):
\begin{align*}
   \frac{\partial \ell}{\partial \beta_{01}} = & Z_{01i}A_{01i}(V_{0i}) \: +  u_{01i}/v_i  & \frac{\partial^2 \ell}{\partial \beta_{01}\partial \beta_{01}} =  & Z_{01i}Z_{01i}^\top A_{01i}(V_{0i}) + \Bigr(v_iu_{0101i} - u_{01i}u_{01i}^\top \Bigr) / v_i^2\\
    \frac{\partial \ell}{\partial \beta_{02}} = & Z_{02i}A_{02i}(V_{0i}) \:  +  u_{02i}/v_i  & \frac{\partial^2 \ell}{\partial \beta_{02}\partial \beta_{02}} =  & Z_{02i}Z_{02i}^\top A_{02i}(V_{0i}) + \Bigr(v_iu_{0202i} - u_{02i}u_{02i}^\top \Bigr) / v_i^2 \\
    \frac{\partial \ell}{\partial \beta_{12}} = & u_{12i}/v_i & \frac{\partial^2 \ell}{\partial \beta_{12}\partial \beta_{12}} =  & \Bigr(v_iu_{1212i} - u_{12i}u_{12i}^\top \Bigr) / v_i^2 \\
    & & \frac{\partial^2 \ell}{\partial \beta_{01}\partial \beta_{02}} = & \Bigr(v_iu_{0102i} - u_{01i}u_{02i}^\top \Bigr) / v_i^2 \\
    & & \frac{\partial^2 \ell}{\partial \beta_{01}\partial \beta_{12}} = & \Bigr(v_iu_{0112i} - u_{01i}u_{12i}^\top \Bigr) / v_i^2 \\
    & & \frac{\partial^2 \ell}{\partial \beta_{02}\partial \beta_{12}} = & \Bigr(v_iu_{0212i} - u_{02i}u_{12i}^\top \Bigr) / v_i^2 
\end{align*}
with, 
\begin{align*}
v_i = & e^{-A_{12i}(T_i)}\int_{L_i}^{R_i}e^{-A_{01i}(u)-A_{02i}(u)+A_{12i}(u)}\alpha_{01i}(u)du \\
u_{01i} = & Z_{01i}\alpha_{12i}(T_i)^{\delta_i^D}e^{-A_{12i}(T_i)} \int_{L_i}^{R_i} \big(1-A_{01i}(u)\big)\alpha_{01i}(u)e^{-A_{01i}(u)-A_{02i}(u)+A_{12i}(u)}du \\ 
u_{02i} = & -Z_{02i}\alpha_{12i}(T_i)^{\delta_i^D}e^{-A_{12i}(T_i)} \int_{L_i}^{R_i} A_{02i}(u)\alpha_{01i}(u)e^{-A_{01i}(u)-A_{02i}(u)+A_{12i}(u)}du \\
u_{12i} = & Z_{12i}e^{-A_{12i}(T_i)}\alpha_{12i}(T_i)^{\delta_i^D} \int_{L_i}^{R_i} \big(A_{12i}(u)- A_{12i}(T_i)+\delta_i^D\big)\alpha_{01i}(u)e^{-A_{01i}(u)-A_{02i}(u)+A_{12i}(u)}du  \\
u_{0101i} = & Z_{01i}Z_{01i}^\top\alpha_{12i}(T_i)^{\delta_i^D}e^{-A_{12i}(T_i)} \int_{L_i}^{R_i} \biggr(\big(1-A_{01i}(u)\big)^2- \\  
& A_{01i}(u)\biggr)\alpha_{01i}(u)e^{-A_{01i}(u)-A_{02i}(u)+A_{12i}(u)}du \\
u_{0202i} = & -Z_{02i}Z_{02i}^\top\alpha_{12i}(T_i)^{\delta_i^D}e^{-A_{12i}(T_i)} \int_{L_i}^{R_i} \big(1-A_{02i}(u)\big)\alpha_{01i}(u)A_{02i}(u)e^{-A_{01i}(u)-A_{02i}(u)+A_{12i}(u)}du \\ \displaybreak
u_{1212i} = & Z_{12i}Z_{12i}^\top\alpha_{12i}(T_i)^{\delta_i^D}e^{-A_{12i}(T_i)}\int_{L_i}^{R_i}
\biggr(\big(A_{12i}(u)-A_{12i}(T_i)\big)^2+A_{12i}(u)-A_{12i}(T_i)+ \\
& \delta_i^D\big(2A_{12i}(u)-2A_{12i}(T_i)+1\big)\biggr)\alpha_{01i}(u)e^{-A_{01i}(u)-A_{02i}(u)+A_{12i}(u)}du \\ 
u_{0102i} = & -Z_{01i}Z_{02i}^\top\alpha_{12i}(T_i)^{\delta_i^D}e^{-A_{12i}(T_i)} \int_{L_i}^{R_i}\big(1-A_{01i}(u)\big)\alpha_{01i}(u)A_{02i}(u)e^{-A_{01i}(u)-A_{02i}(u)+A_{12i}(u)}du \\
u_{0112i} = & Z_{01i}Z_{12i}^\top\alpha_{12i}(T_i)^{\delta_i^D}e^{-A_{12i}(T_i)} \int_{L_i}^{R_i}\big(1-A_{01i}(u)\big)\big(A_{12i}(u)- \\
& A_{12i}(T_i)+1^{\delta_i^D}\big)\alpha_{01i}(u)e^{-A_{01i}(u)-A_{02i}(u)+A_{12i}(u)}du \\
u_{0212i} = & -Z_{02i}Z_{12i}^\top\alpha_{12i}(T_i)^{\delta_i^D}e^{-A_{12i}(T_i)} \int_{L_i}^{R_i}A_{02i}(u)\big(A_{12i}(u)-\\
& A_{12i}(T_i)+1^{\delta_i^D}\big)\alpha_{01i}(u)e^{-A_{01i}(u)-A_{02i}(u)+A_{12i}(u)}du \\
\end{align*}

For individual i, who entered the study at $V_{0i}$, remained healthy until visit $L_i$ and \\ was either censored at time $T_i$ (case 3) or dead at $T_i$ (case 4):
\begin{align*}
   \frac{\partial \ell}{\partial \beta_{01}} = & Z_{01i}A_{01i}(V_{0i}) \: +  u_{01i}/v_i  & \frac{\partial^2 \ell}{\partial \beta_{01}\partial \beta_{01}} =  & Z_{01i}Z_{01i}^\top A_{01i}(V_{0i}) + \Bigr(v_iu_{0101i} - u_{01i}u_{01i}^\top \Bigr) / v_i^2\\
    \frac{\partial \ell}{\partial \beta_{02}} = & Z_{02i}A_{02i}(V_{0i}) \:  +  u_{02i}/v_i  & \frac{\partial^2 \ell}{\partial \beta_{02}\partial \beta_{02}} =  & Z_{02i}Z_{02i}^\top A_{02i}(V_{0i}) + \Bigr(v_iu_{0202i} - u_{02i}u_{02i}^\top \Bigr) / v_i^2 \\
    \frac{\partial \ell}{\partial \beta_{12}} = & u_{12i}/v_i & \frac{\partial^2 \ell}{\partial \beta_{12} \partial \beta_{12}} =  & \Bigr(v_iu_{1212i} - u_{12i}u_{12i}^\top \Bigr) / v_i^2 \\
    & & \frac{\partial^2 \ell}{\partial \beta_{01}\partial \beta_{02}} =  & \Bigr(v_iu_{0102i} - u_{01i}u_{02i}^\top \Bigr) / v_i^2\\
     & & \frac{\partial^2 \ell}{\partial \beta_{01}\partial \beta_{12}} =  & \Bigr(v_iu_{0112i} - u_{01i}u_{12i}^\top \Bigr) / v_i^2\\
      & & \frac{\partial^2 \ell}{\partial \beta_{02}\partial \beta_{12}} =  & \Bigr(v_iu_{0212i} - u_{02i}u_{12i}^\top \Bigr) / v_i^2
\end{align*}
with, 
\begin{align*}
v_i = &  e^{-A_{01i}(T_i)-A_{02i}(T_i)}\alpha_{02i}(T_i)^{\delta_i^D}+e^{-A_{12i}(T_i)}\int_{L_i}^{T_i}e^{-A_{01i}(u)-A_{02i}(u)+A_{12i}(u)}\alpha_{01i}(u)du \\
u_{01i} = & -Z_{01i}A_{01i}(T_i)e^{-A_{01i}(T_i)-A_{02i}(T_i)}\alpha_{02i}(T_i)^{\delta_i^D} + \\
& Z_{01i}\alpha_{12i}(T_i)^{\delta_i^D}e^{-A_{12i}(T_i)} \int_{L_i}^{T_i} \big(-A_{01i}(u)+1\big)\alpha_{01i}(u)e^{-A_{01i}(u)-A_{02i}(u)+A_{12i}(u)}du \\ \displaybreak
u_{02i} = & -Z_{02i}(A_{02i}(T_i)-\delta_i^D)e^{-A_{01i}(T_i)-A_{02i}(T_i)}\alpha_{02i}(T_i)^{\delta_i^D}- \\ & Z_{02i}\alpha_{12i}(T_i)^{\delta_i^D}e^{-A_{12i}(T_i)} \int_{L_i}^{T_i} A_{02i}(u)\alpha_{01i}(u)e^{-A_{01i}(u)-A_{02i}(u)+A_{12i}(u)}du \\
u_{12i} = & Z_{12i}e^{-A_{12i}(T_i)}\alpha_{12i}(T_i)^{\delta_i^D} \int_{L_i}^{T_i} \big(A_{12i}(u)-A_{12i}(T_i)+\delta_i^D\big)\alpha_{01i}(u)e^{-A_{01i}(u)-A_{02i}(u)+A_{12i}(u)}du  \\
u_{0101i} = & -Z_{01i}Z_{01i}^\top A_{01i}(T_i)e^{-A_{01i}(T_i)-A_{02i}(T_i)}(1-A_{01i}(T_i))\alpha_{02i}(T_i)^{\delta_i^D}+ \\ & Z_{01i}Z_{01i}^\top\alpha_{12i}(T_i)^{\delta_i^D}e^{-A_{12i}(T_i)} \int_{L_i}^{T_i} \biggr(\big(1-A_{01i}(u)\big)^2 - \\
& -A_{01i}(u)\biggr)\alpha_{01i}(u)e^{-A_{01i}(u)-A_{02i}(u)+A_{12i}(u)}du \\
u_{0202i} = & -Z_{02i}Z_{02i}^\top A_{02i}(T_i)e^{-A_{01i}(T_i)-A_{02i}(T_i)}\biggr(1-A_{02i}(T_i)+ \\
& \delta_i^D\bigr(\big(1-A_{02i}(T_i))^2+2(1-A_{02i}(T_i)\big)\bigr)\biggr)\alpha_{02i}(T_i)^{\delta_i^D}- \\
& Z_{02i}Z_{02i}^\top\alpha_{12i}(T_i)^{\delta_i^D}e^{-A_{12i}(T_i)} \int_{L_i}^{T_i} \big(1-A_{02i}(u)\big)\alpha_{01i}(u)A_{02i}(u)e^{-A_{01i}(u)-A_{02i}(u)+A_{12i}(u)}du \\
u_{1212i} = & Z_{12i}Z_{12i}^\top\alpha_{12i}(T_i)^{\delta_i^D}e^{-A_{12i}(T_i)}\int_{L_i}^{T_i}
\biggr(\big(A_{12i}(u)-A_{12i}(T_i)\big)^2+A_{12i}(u)-A_{12i}(T_i)+ \\
& \delta_i^D\big(2A_{12i}(u)-2A_{12i}(T_i)+1\big)\biggr)\alpha_{01i}(u)e^{-A_{01i}(u)-A_{02i}(u)+A_{12i}(u)}du \\
u_{0102i} = & Z_{01i}Z_{02i}^\top A_{01i}(T_i)\big(A_{02i}(T_i)-1^{\delta_i^D}\big)\alpha_{02i}(T_i)^{\delta_i^D}e^{-A_{01i}(T_i)-A_{02i}(T_i)} -\\ & Z_{01i}Z_{02i}^\top\alpha_{12i}(T_i)^{\delta_i^D}e^{-A_{12i}(T_i)} \int_{L_i}^{T_i} \big(1-A_{01i}(u)\big)\alpha_{01i}(u)A_{02i}(u)e^{-A_{01i}(u)-A_{02i}(u)+A_{12i}(u)}du \\
u_{0112i} =  & Z_{01i}Z_{12i}^\top\alpha_{12i}(T_i)^{\delta_i^D}e^{-A_{12i}(T_i)} \int_{L_i}^{T_i} \big(1-A_{01i}(u)\big)\big(A_{12i}(u)- \\
& A_{12i}(T_i)+ 1^{\delta_i^D}\big)\alpha_{01i}(u)e^{-A_{01i}(u)-A_{02i}(u)+A_{12i}(u)}du \\
u_{0212i} =  & -Z_{02i}Z_{12i}^\top\alpha_{12i}(T_i)^{\delta_i^D}e^{-A_{12i}(T_i)} \int_{L_i}^{T_i}A_{02i}(u)(A_{12i}(u) - \\
& A_{12i}(T_i)+1^{\delta_i^D}\big)\alpha_{01i}(u)e^{-A_{01i}(u)-A_{02i}(u)+A_{12i}(u)}du \\
\end{align*}

\newpage

\begin{table}
\centering
\caption{Simulated covariate parameters on transitions from {\em healthy} to {\em diseased} ($0 \rightarrow 1$), {\em diseased} to {\em deceased} ($1 \rightarrow 2$), and {\em healthy} to {\em deceased} ($0 \rightarrow 2$).}
\label{covariate_effect}
\begin{tabular}{cccccccccccccc}  
\hline
Transition & $\beta_1$ & $\beta_2$ & $\beta_3$ & $\beta_4$ & $\ldots$ & $\beta_{10}$ & $\beta_{11}$ & $\beta_{12}$ & $\beta_{13}$ & $\beta_{14}$ & $\ldots$ & $\beta_{20}$ & \\ \hline 
$0 \rightarrow 1$   & 0.8 & 0.8 & 0.8 & 0 & $\ldots$ & 0 & -0.8 & -0.5 & -0.5 & 0 & $\ldots$ & 0 & \\
$0 \rightarrow 2$   & 0 & 0.8 & 0.8 & 0 & $\ldots$ & 0 & 0 & 0 & 0.5 & 0 & $\ldots$ & 0 & \\
$1 \rightarrow 2$   & 0 & 0.8 & 0 & 0 & $\ldots$ & 0 & 0 & -0.5 & 0 & 0 & $\ldots$ & 0 & \\ \hline

& $\beta_{21}$ & $\beta_{22}$ & $\beta_{23}$ & $\beta_{24}$ & $\ldots$ & $\beta_{30}$ & $\beta_{31}$ & $\beta_{32}$ & $\beta_{33}$ & $\beta_{34}$ & $\ldots$ & $\beta_{40}$ & \\ \cline{2-14}
$0 \rightarrow 1$   & 0 & 0 & 0 & 0 & $\ldots$ & 0 & 0 & 0 & 0 & 0 & $\ldots$ & 0 & \\
$0 \rightarrow 2$   & -0.8 & -0.5 & 0 & 0 & $\ldots$ & 0 & 0.8 & 0.8 & 0 & 0 & $\ldots$ & 0 & \\
$1 \rightarrow 2$   & 0 & -0.5 & -0.8 & 0 & $\ldots$ & 0 & 0 & 0.8 & 0.8 & 0 & $\ldots$ & 0 & \\ \hline

& $\beta_{41}$ & $\beta_{42}$ & $\beta_{43}$ & $\beta_{44}$ & $\ldots$ & $\beta_{50}$   \\ \cline{2-7}
$0 \rightarrow 1$   & 0 & -0.5 & 0 & 0 & $\ldots$ & 0 \\
$0 \rightarrow 2$   & 0 & 0.5 & 0.5 & 0 & $\ldots$ & 0   \\
$1 \rightarrow 2$   & -0.5 & -0.5 & 0 & 0 & $\ldots$ & 0   \\ \hline

\end{tabular}
\end{table}

\begin{sidewaystable}
    \centering
     \caption{Description of the six simulated scenarios based on events rates, sparsity between visits and under two correlation structure between covariates. \\ }
      \label{scenarios_descript}
       \resizebox{1\textwidth}{0.18\textheight}{
    \begin{tabular}{lcccccc}
\hline
\multicolumn{1}{c}{Scenario} & \multicolumn{3}{c}{Transitions, N( \%)\textsuperscript{*}} & Undiagnosed disease & \multicolumn{2}{c}{Weibull parameters\textsuperscript{***}}  \\ 
& $0\rightarrow1$ & $0\rightarrow2$ & $1\rightarrow2$ & because of death, N( \%)\textsuperscript{**} & $\theta_1$ & $\theta_2$ \\ \hline \\
\multicolumn{5}{l}{\bf Scenario A: low event rates and frequent visit} & (2,2.5,2.5) & (0.015,0.025,0.04)\\
{\bf(1)} Independent covariates \\
\multicolumn{1}{r}{True} & 288 (14) & 480 (24) & 133 (7) \\
\multicolumn{1}{r}{Observed} & 197 (10) & 535 (27) & 79 (4) & 31 (11) \\
{\bf(2)} group-Toeplitz covariates \\
\multicolumn{1}{r}{True} & 360 (18) & 506 (25) & 190 (10) \\
\multicolumn{1}{r}{Observed} & 248 (12) & 581 (29) & 115 (6) & 45 (13)\\ \\
\multicolumn{5}{l}{\bf Scenario B: high event rates and frequent visit}& (3.4,3.4,3.4) & (0.075,0.085,0.07)\\
{\bf(3)} Independent covariates  \\
\multicolumn{1}{r}{True} & 825 (41) & 1,051 (53) & 574 (29) \\
\multicolumn{1}{r}{Observed} & 565 (28) & 1,259 (63) & 365 (18) & 118 (14) \\
{\bf(4)} group-Toeplitz covariates \\
\multicolumn{1}{r}{True} & 816 (41) & 1,006 (50) & 565 (28) \\
\multicolumn{1}{r}{Observed} & 571 (29) & 1,200 (60) & 370 (19) & 113 (14)  \\ \\
 \multicolumn{5}{l}{\bf Scenario C: high event rates and sparse visit} & (3.4,3.4,3.4) & (0.08,0.08,0.08) \\
{\bf(5)} Independent covariates \\
\multicolumn{1}{r}{True} & 934 (47) & 943 (47) & 714 (36) \\
\multicolumn{1}{r}{Observed} & 528 (26) & 1,280 (64) & 376 (19) & 230 (25) \\
{\bf(6)} group-Toeplitz covariates \\
\multicolumn{1}{r}{True} & 910 (46) & 912 (46) & 683 (34) \\
\multicolumn{1}{r}{Observed} & 526 (26) & 1,226 (61) & 369 (18) & 222 (24)  \\ \hline
\multicolumn{7}{p{0.95\textwidth}}{\small{\textsuperscript{*}Mean count and percentage over the 500 replicates, each composed of 2,000 subjects. }}\\
\multicolumn{7}{p{0.95\textwidth}}{\small{\textsuperscript{**}Mean count and percentage over the 500 replicates, each composed only of the truly ill subjects. }}\\
\multicolumn{7}{p{0.95\textwidth}}{\small{\textsuperscript{***} with Weibull baseline intensity transition $\alpha_{0}(t|\theta_{1},\theta_{2})=\theta_{1}\theta_{2}(\theta_{2}t)^{\theta_{1}-1}$ and transitions ($0\rightarrow1$,$0\rightarrow2$,$1\rightarrow2$).}}\\
\end{tabular}}

\end{sidewaystable}

\begin{sidewaystable}
\centering
 \caption{ Distribution of True Positive Rate (TPR) and False Positive Rate (FPR) using Reg IDM-TT, Reg IDM-ICT and Reg PHM with M-spline basis for baseline transition, under different simulation scenario settings (over 500 replicates).
    }
    \label{TPR-FPR}
    \resizebox{1\textwidth}{0.18\textheight}{
\begin{tabular}{lccccccccccccccccc}
       \hline    & \multicolumn{8}{c}{Independent} & &  \multicolumn{8}{c}{group-Toeplitz} \\        
        & \multicolumn{2}{c}{Scenario A} & & \multicolumn{2}{c}{Scenario B} & &\multicolumn{2}{c}{Scenario C} & & \multicolumn{2}{c}{Scenario A} & &  \multicolumn{2}{c}{Scenario B} & & \multicolumn{2}{c}{Scenario C} \\ 
        \cline{2-3}  \cline{5-6}  \cline{8-9}  \cline{11-12} \cline{14-15} \cline{17-18} & TPR & FPR & & TPR & FPR & & TPR & FPR & & TPR & FPR & & TPR & FPR & & TPR & FPR \\ \hline
        
 \textbf{Transition $0 \rightarrow 1$}      \\ 
Reg IDM-TT & 1.00  (0.00) & 0.05  (0.04) & & 1.00 (0.00) & 0.06 (0.05) & & 1.00 (0.00) & 0.07 (0.05) & & 1.00 (0.01) & 0.00 (0.00) & & 1.00 (0.00) & 0.04 (0.04) & & 1.00 (0.00) & 0.04 (0.04)\\ 
Reg IDM-ICT & 1.00 (0.00) & 0.05 (0.04) & & 1.00 (0.00) & 0.06 (0.05) & & 1.00 (0.00) & 0.04 (0.04) & & 0.98 (0.05)  &  0.00 (0.01) & & 1.00 (0.00) & 0.04 (0.04) & & 1.00 (0.00) & 0.03 (0.04)\\
Reg PHM & 1.00 (0.01) & 0.06 (0.05) & & 1.00 (0.00) & 0.12 (0.06) & & 1.00 (0.00) & 0.15 (0.06) & & 0.95 (0.07)  & 0.00 (0.01) & & 1.00 (0.00) & 0.09 (0.04) & &  0.99 (0.04) & 0.07 (0.04) \\ \\
 \textbf{Transition $0 \rightarrow 2$}      \\ 
Reg IDM-TT & 1.00 (0.00) &  0.08 (0.06) & & 1.00 (0.00) & 0.10 (0.06) & & 1.00 (0.00) & 0.09 (0.06) & & 1.00 (0.00)  & 0.03 (0.03) & &1.00 (0.00) & 0.06 (0.04) & & 1.00 (0.00) &  0.06 (0.04)\\ 
Reg IDM-ICT & 1.00 (0.00) & 0.08 (0.06) & & 1.00 (0.00) & 0.09 (0.06) & & 1.00 (0.00) & 0.10 (0.07) & & 1.00 (0.00) & 0.04 (0.04) & & 1.00 (0.00) & 0.05 (0.05) & & 1.00 (0.00) &  0.05 (0.05) \\
Reg PHM & 1.00 (0.00) & 0.09 (0.06) & & 1.00 (0.00) &  0.16 (0.08) & & 1.00 (0.00) & 0.21 (0.08) & & 1.00 (0.01) & 0.06 (0.05) & & 1.00 (0.00) & 0.09 (0.05) & & 1.00 (0.01) &  0.12 (0.06)\\ \\
 \textbf{Transition $1 \rightarrow 2$}      \\ 
Reg IDM-TT & 0.99 (0.03) & 0.08 (0.05) & & 1.00 (0.00) & 0.10 (0.06) & & 1.00 (0.00) & 0.09 (0.06) & &  0.98 (0.05) & 0.05 (0.04) & & 1.00 (0.01) & 0.06 (0.05) & & 1.00 (0.01) & 0.05 (0.04) \\ 
Reg IDM-ICT & 0.92 (0.17)  & 0.06 (0.05) & & 1.00 (0.00) & 0.10 (0.06) & & 1.00 (0.00) & 0.09 (0.06) & & 0.85 (0.14) & 0.02 (0.03) & & 1.00 (0.01) & 0.04 (0.05) & & 1.00 (0.01) & 0.05 (0.05)\\  \hline 
 \multicolumn{12}{l}{\small{Mean (Standard Deviation)}} \\

 \multicolumn{12}{l}{\small{Scenarios (A) low event rates and frequent visit schedule, (B) high event rates and frequent visit schedule and (C) high event rates and sparse visit schedule. }}\\
 \multicolumn{12}{l}{\small{\textit{Abbreviations:} TPR=True Positive Rate ($|\mathcal I(\widehat \beta \ne 0) \in \mathcal I(\beta \ne 0)|$/$|I(\beta \ne 0)|$), FPR=False Positive Rate ($|\mathcal I(\widehat \beta \ne 0) \in \mathcal I(\beta = 0)|$/$|I(\beta = 0)|$), }}\\
  \multicolumn{12}{l}{\small{IC=Interval-Censoring, IDM=Illness-Death Model, ICT=Interval-Censored Time, Reg=Regularized, TT=True Times, PHM=Proportional Hazard Model.}}
\end{tabular}}
\end{sidewaystable}

\begin{sidewaystable}
\centering
 \caption{Characteristics of the 1984 subjects of 3C sample according to their diagnosis of dementia, life status and overall at baseline 
    }\label{descrip3C}
 \resizebox{1\textwidth}{0.18\textheight}{\begin{tabular}{lccccc}
\hline
& & \multicolumn{2}{c}{\textbf{Alive}} & \multicolumn{2}{c}{\textbf{Dead}} \\
\textbf{Characteristic*} & \textbf{Overall} & \textbf{Dementia-free} & \textbf{With dementia} & \textbf{Dementia-free} & \textbf{With dementia}\\
& N = 1,979 &  N = 1,378 &    N = 109 &  N = 403 &  N = 89\\
\hline \\
\textbf{Socio-demographic}\\
Female sex (n, \%) &  1,216 (61) & 901 (65) & 81 (74) & 181 (45) & 53 (60)\\
Short educational level (n, ) &  1,166 (59) & 808 (59) & 64 (59) & 245 (61) & 49 (55)\\
Age (in years) &  72.27 (4.00) & 71.63 (3.81) & 73.23 (3.75) & 73.57 (4.06) & 75.28 (3.79)\\ \\
\textbf{Health examination} \\
Systolic blood pressure (mmHg) &   146.89 (22.53) & 146.55 (22.22) & 145.93 (21.04) & 148.60 (23.43) & 145.42 (24.83)\\
Diastolic blood pressure (mmHg) &  83.71 (11.46) & 84.21 (11.26) & 82.71 (11.07) & 82.89 (11.83) & 80.99 (12.87)\\
History of coronary disease (n, \%)&  93 (4.7) & 49 (3.6) & 7 (6.4) & 32 (7.9) & 5 (5.6) \\
HBP (n, \%)& 755 (38) & 489 (35) & 43 (39) & 189 (47) & 34 (38)\\
BMI (in kg/m2) &  25.52 (3.73) & 25.36 (3.68) & 25.66 (3.30) & 26.00 (3.81) & 25.60 (4.55) \\
CESD & 9.05 (8.45) & 9.28 (8.45) & 9.63 (9.59) & 8.20 (8.01) & 8.63 (8.61)\\
History of diabetes (n, \%) &  148 (7.5) & 85 (6.2) & 11 (10) & 44 (11) & 8 (9.0)\\
Number of drugs used (/month) &  4.04 (2.82) & 3.85 (2.71) & 4.22 (2.74) & 4.43 (3.08) & 4.97 (3.08)\\
Dependent in IADL (n, \%) & 118 (6.0) & 77 (5.6) & 7 (6.4) & 25 (6.2) & 9 (10)\\
Dependent in mobility (n, \%)& 76 (3.8) & 43 (3.1) & 6 (5.5) & 22 (5.5) & 5 (5.6) \\ \\
\textbf{Cognition} \\
MMSE &  27.80 (1.70) & 27.87 (1.70) & 27.44 (1.86) & 27.78 (1.64) & 27.21 (1.72)\\
Benton Visual Retention &  11.79 (1.84) & 11.96 (1.75) & 11.17 (2.13) & 11.55 (1.93) & 10.81 (1.78) \\
Isaacs’s Set Test &  33.54 (6.68) & 34.42 (6.50) & 30.87 (6.37) & 32.18 (6.85) & 29.44 (5.53)\\
TMT A (\# good moves /time) & 30.29 (9.83) & 31.21 (9.64) & 27.41 (8.84) & 29.31 (10.40) & 24.11 (7.95)\\
TMT Test B (\# good moves /time) &  14.15 (7.51) & 14.87 (7.76) & 11.98 (6.01) & 13.17 (6.91) & 10.13 (5.42)\\ \\
\end{tabular}}
\end{sidewaystable}

\clearpage

\begin{sidewaystable}

 \resizebox{1\textwidth}{0.18\textheight}{\begin{tabular}{lccccc}
\hline
& & \multicolumn{2}{c}{\textbf{Alive}} & \multicolumn{2}{c}{\textbf{Dead}} \\
\textbf{Characteristic*} & \textbf{Overall} & \textbf{Dementia-free} & \textbf{With dementia} & \textbf{Dementia-free} & \textbf{With dementia}\\
& N = 1,979 &  N = 1,378 &    N = 109 &  N = 403 &  N = 89\\
\hline \\
\textbf{Genome} \\
At least one ApoE $\epsilon$4 allele (n, \%) &  430 (22) & 289 (21) & 33 (30) & 79 (20) & 29 (33)\\ \\
\textbf{MRI} \\
TIV (in cm3) &   1,313.29 (147.87) & 1,326.34 (144.99) & 1,252.04 (137.28) & 1,297.53 (155.26) & 1,257.58 (133.08)\\
Grey Matter Volume (in cm3) & 499.98 (49.65) & 504.06 (49.56) & 483.65 (41.89) & 494.79 (50.58) & 480.36 (44.99)\\
WM Volume (in cm3) &  465.98 (52.39) & 466.04 (52.80) & 453.23 (45.86) & 470.48 (52.32) & 460.19 (51.55)\\
Hippocampus volume (in cm3) &  6.63 (0.81) & 6.70 (0.78) & 6.33 (0.81) & 6.55 (0.82) & 6.23 (0.88)\\
Median temporal lobe (in cm3) &  15.67 (1.73) & 15.80 (1.68) & 15.05 (1.86) & 15.58 (1.75) & 14.83 (1.88)\\
Deep WMH (in \% of WM mask) &  0.70 (0.51) & 0.64 (0.47) & 0.74 (0.49) & 0.86 (0.58) & 0.90 (0.61)\\
Perivascular WMH (in \% of WM mask) &  1.49 (1.54) & 1.43 (1.44) & 1.48 (1.20) & 1.56 (1.43) & 1.97 (3.05)\\
WM PVS (\# of)&  303.75 (109.73) & 298.79 (111.43) & 319.66 (111.11) & 314.17 (105.20) & 313.97 (96.61)\\
Basal Ganglia PVS (\# of) & 19.22 (8.71) & 18.34 (8.17) & 19.10 (8.20) & 21.78 (9.99) & 21.38 (8.66)\\ \hline
\multicolumn{6}{l}{\small{* Mean (Standard Deviation) for continuous data and Sample size (percentage) for binary data} }\\
\multicolumn{6}{l}{\small \textit{Abbreviations: }N=sample size, HBP=High Blood Pressure, BMI=Body Mass Index, CESD = Center for Epidemiologic Studies Depression Scale, }\\
\multicolumn{6}{l}{\small IADL=Instrumental Activities of Daily Living, MMSE=Mini-Mental State Examination, TMT=Trail Making Test, TIV=Total Intracranial Volume,}\\
\multicolumn{6}{l}{\small WM=White matter, GM=grey matter, WMH= White matter hyperintensities and PVS=PeriVascular Spaces.}\\
\end{tabular}}
\end{sidewaystable}

\begin{figure}
    \centering
    \includegraphics[width=16cm]{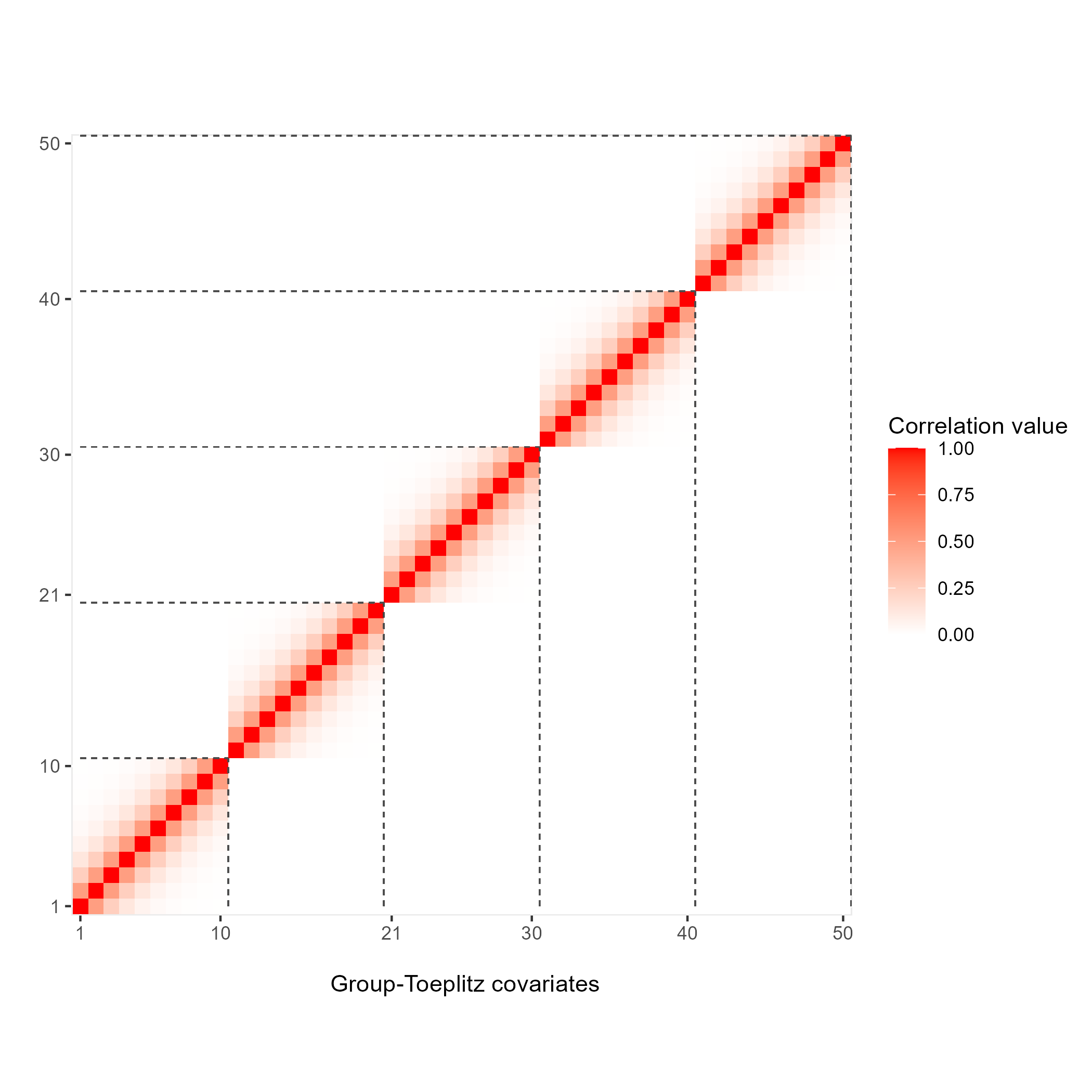}
    \caption{Correlation heatmap for the fifty group-Toeplitz covariates in simulation.}
    \label{heatmap}
\end{figure}

\begin{figure}
    \centering
    \hspace*{-2cm}\includegraphics[width=18cm,height=18cm]{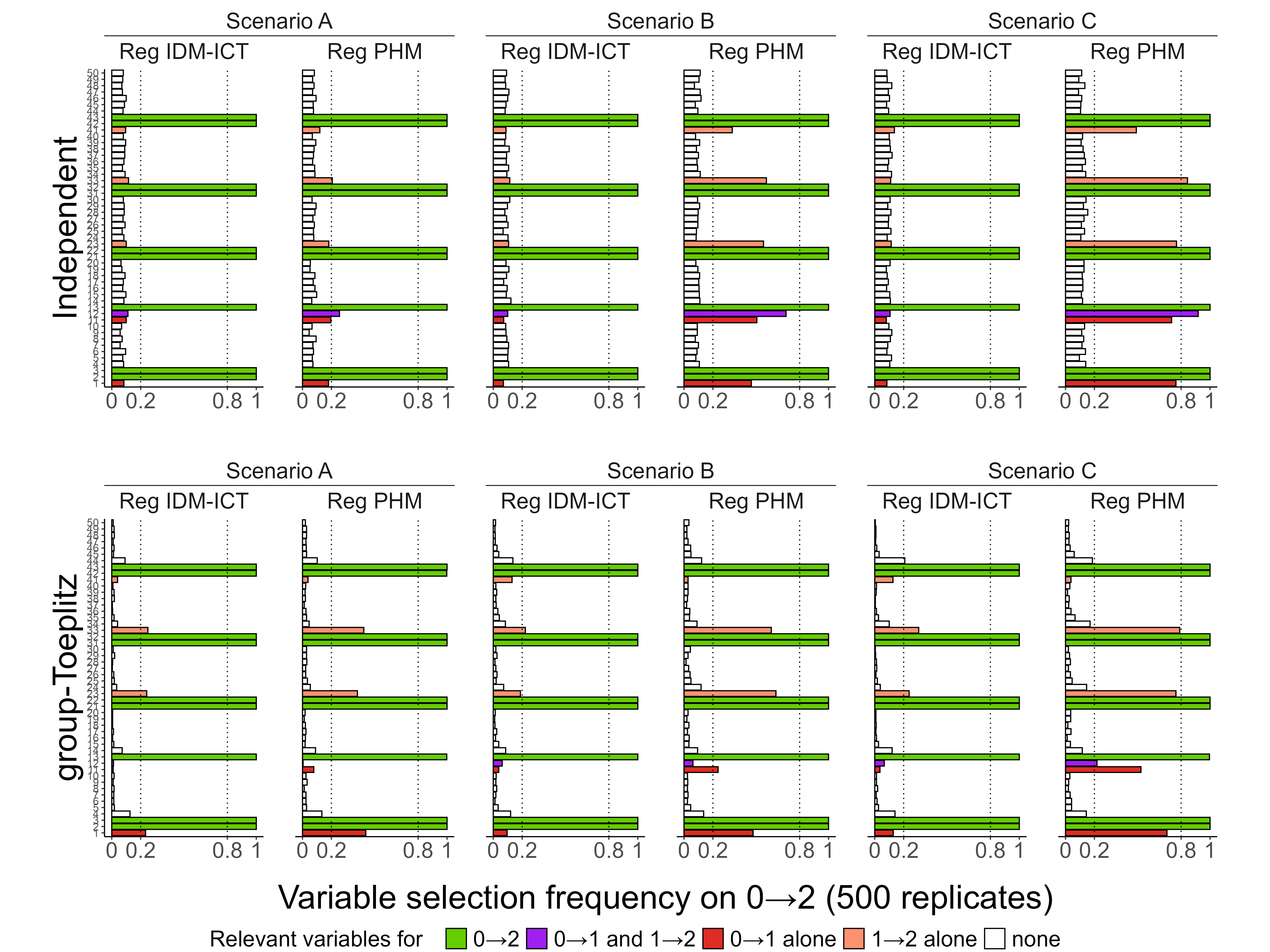}
    \caption{Variable selection frequency over 500 simulated replicates for the transition  {\em healthy} to {\em death}, using Reg IDM-ICT and Reg PHM  with M-spline basis for baseline transition, under low event rates with a frequent visit schedule (A) or high event rates with a frequent or sparse visit schedule (B and C) and two correlation structures among the covariates (Independent or group-Toeplitz).}
    \label{selection_02}
\end{figure}

\begin{figure}
    \centering
    \hspace*{-2cm}\includegraphics[width=18cm,height=18cm]{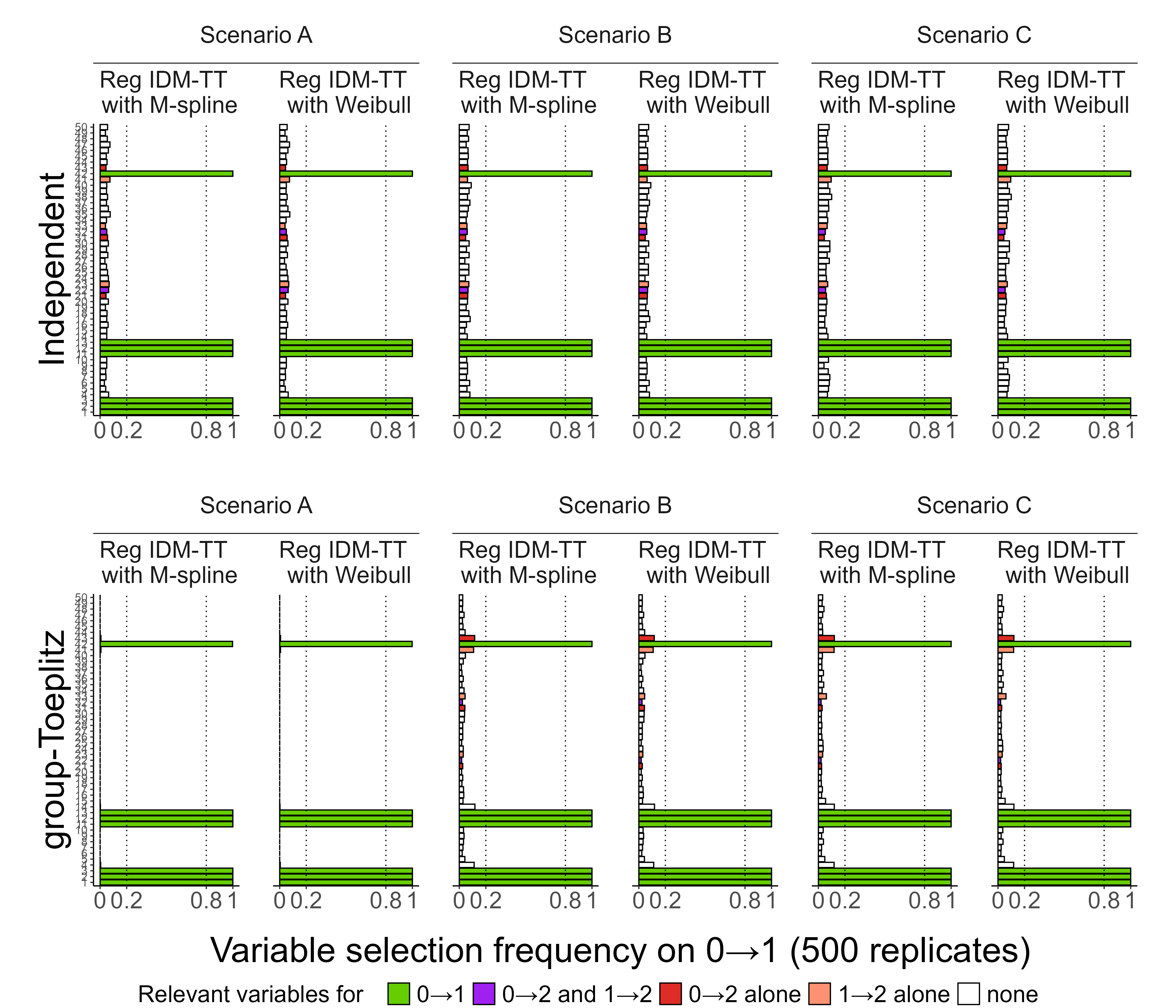}
    \caption{ Variable selection frequency over 500 simulated replicates for the transition {\em healthy} to {\em illness}, using Reg IDM-TT with M-spline basis or Weibull for baseline transition, under low event rates with a frequent visit schedule (A) or high event rates with a frequent or sparse visit schedule (B and C) and two correlation structures among the covariates (Independent or group-Toeplitz).}
    \label{selection_01_true}
\end{figure}

\begin{figure}
    \centering
   \hspace*{-2cm}\includegraphics[width=18cm,height=18cm]{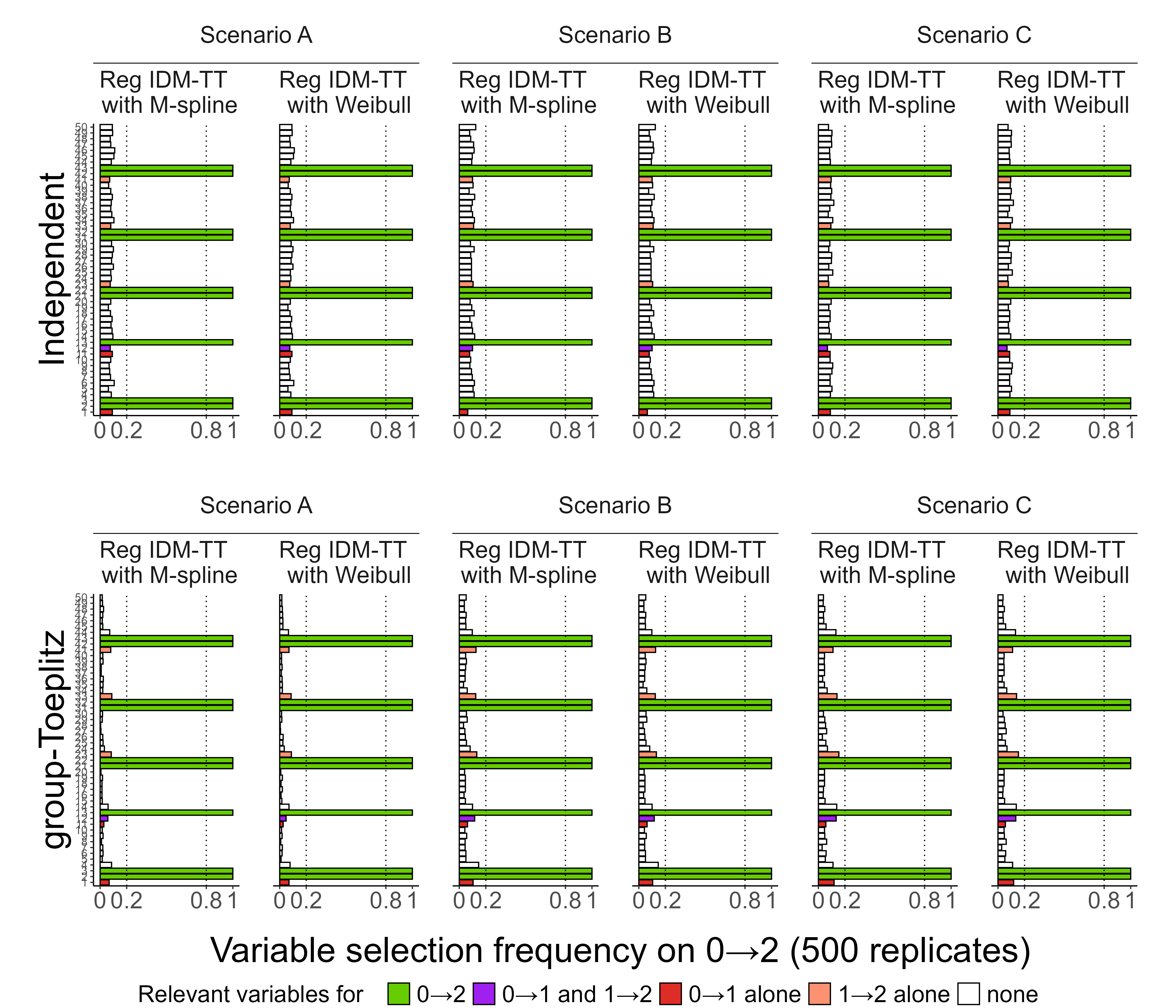}
    \caption{Variable selection frequency over 500 simulated replicates for the transition {\em healthy} to {\em death}, using Reg IDM-TT with M-spline basis or Weibull for baseline transition, under low event rates with a frequent visit schedule (A) or high event rates with a frequent or sparse visit schedule (B and C) and two correlation structures among the covariates (Independent or group-Toeplitz).}
    \label{selection_02_true}
\end{figure}

\begin{figure}
    \centering
    \hspace*{-2cm}\includegraphics[width=18cm,height=18cm]{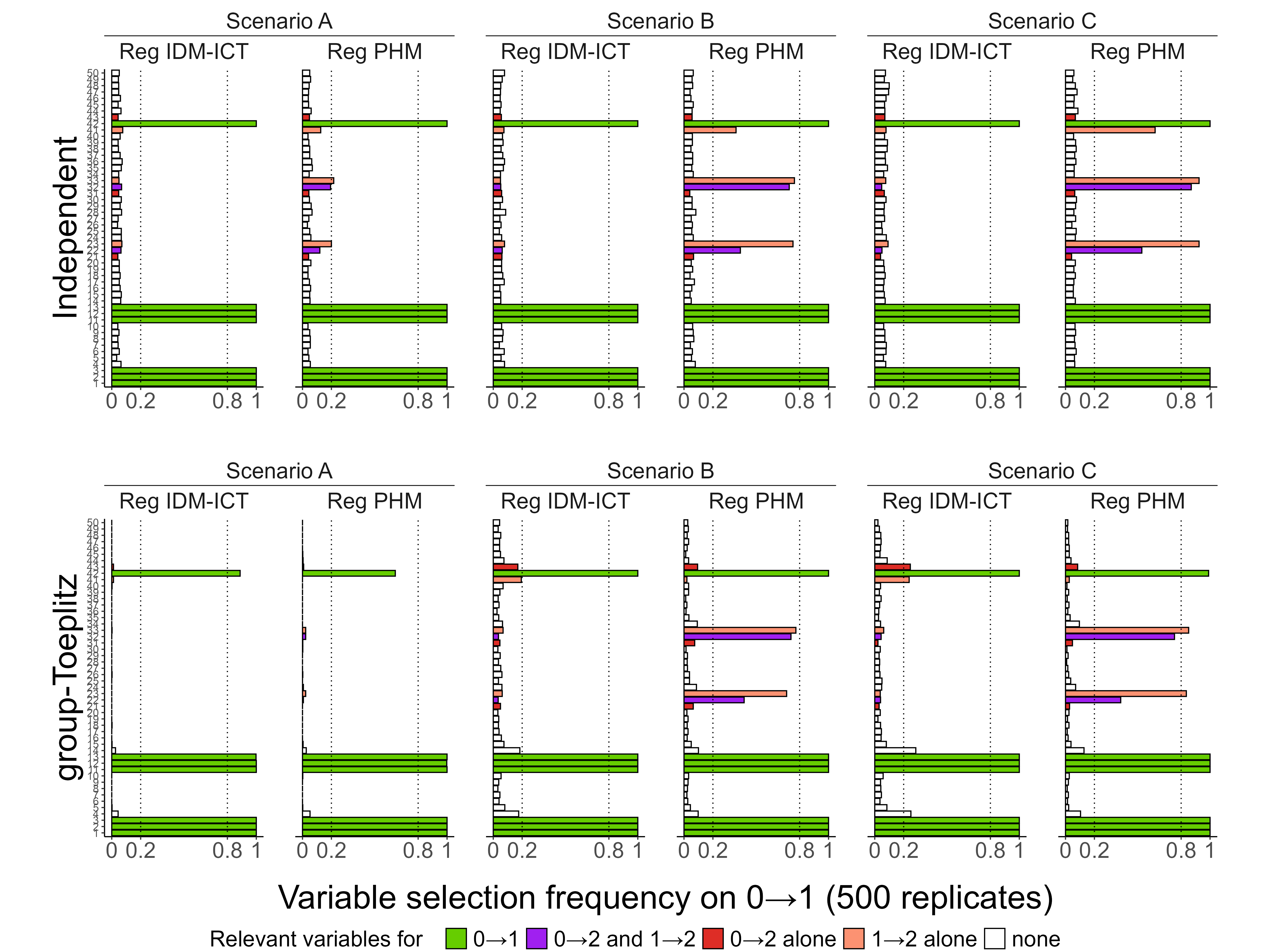}
    \caption{Variable selection frequency over 500 simulated replicates for the transition  {\em healthy} to {\em illness}, using Reg IDM-ICT and Reg PHM with Weibull for baseline transition, under low event rates with a frequent visit schedule (A) or high event rates with a frequent or sparse visit schedule (B and C) and two correlation structures among the covariates (Independent or group-Toeplitz).}
    \label{selection_weib_01}
\end{figure}

\begin{figure}
    \centering
    \hspace*{-2cm}\includegraphics[width=18cm,height=18cm]{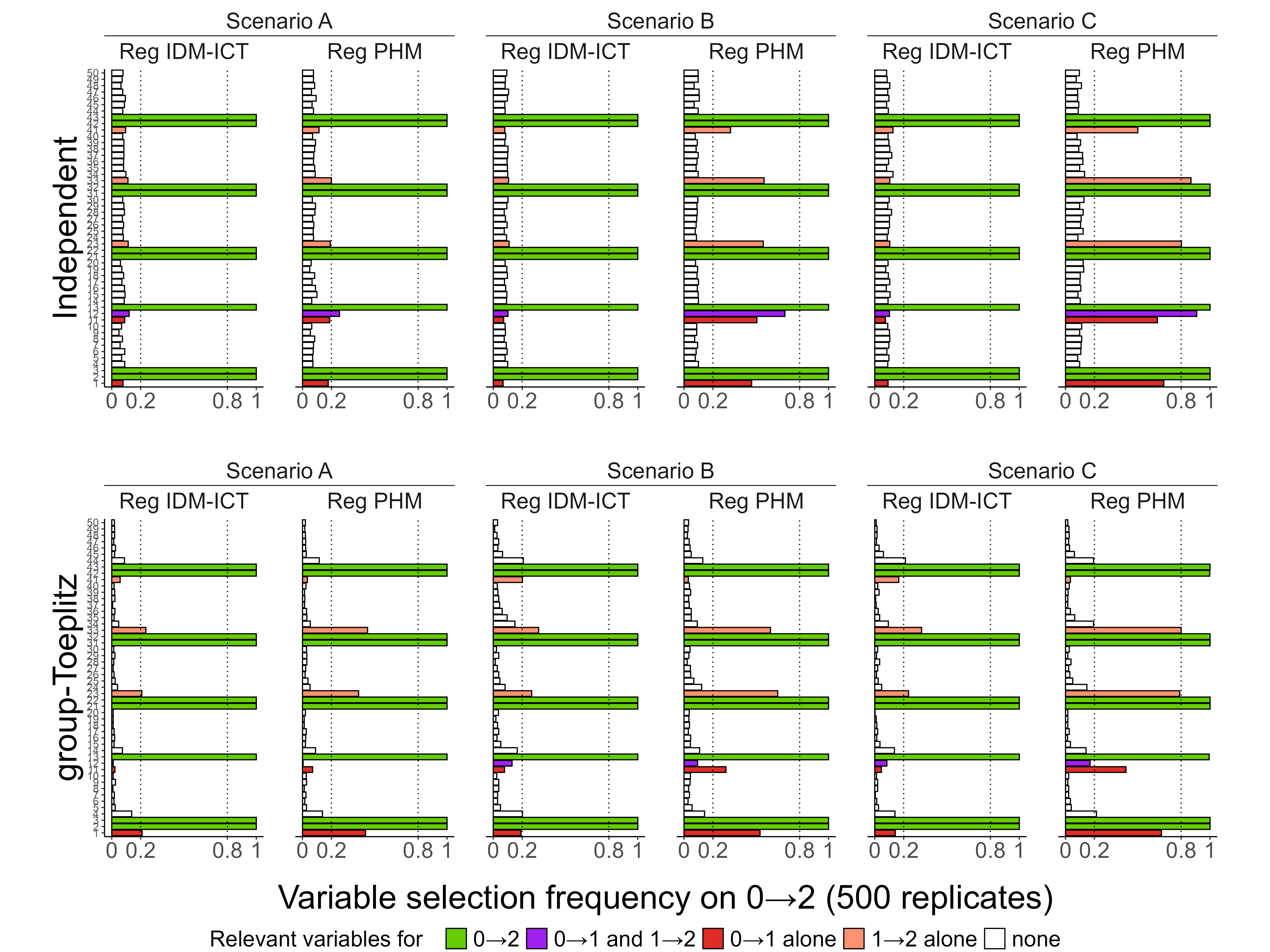}
    \caption{Variable selection frequency over 500 simulated replicates for the transition  {\em healthy} to {\em death}, using Reg IDM-ICT and Reg PHM with Weibull for baseline transition, under low event rates with a frequent visit schedule (A) or high event rates with a frequent or sparse visit schedule (B and C) and two correlation structures among the covariates (Independent or group-Toeplitz).}
    \label{selection_weib_02}
\end{figure}

\begin{figure}
    \centering
    \hspace*{-12ex}\includegraphics[width=18cm]{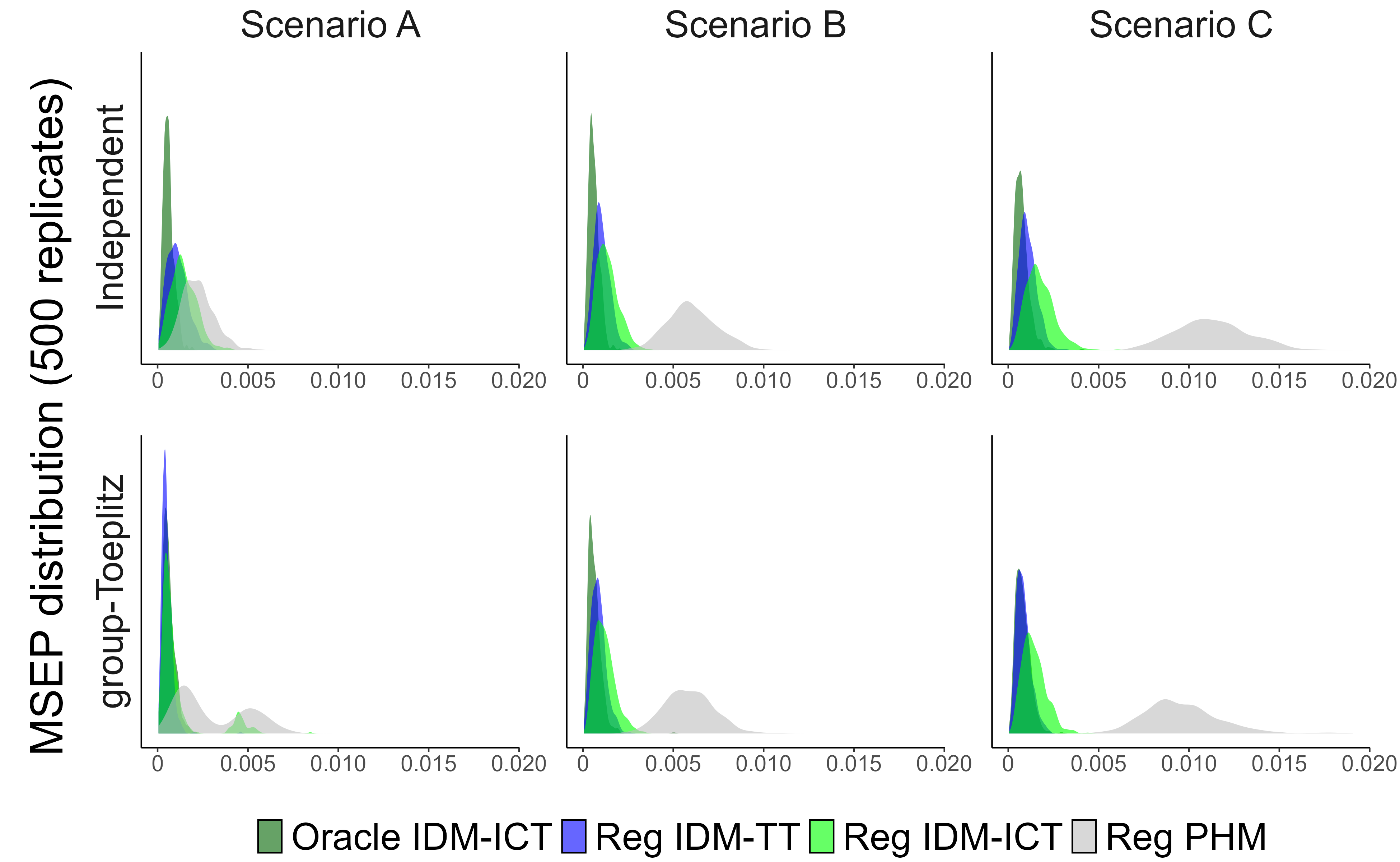}
    \caption{Distribution over 500 replicates of the Mean Square Error of the Probability (MSEP) for a healthy subject to become ill, using the Oracle IDM-ICT ({\color{green(ryb)}dark green}), the Reg IDM-TT ({\color{blue}blue}), Reg IDM-ICT ({\color{green(light)}green}) and the Reg PHM ({\color{lightgray}grey}) with Weibull for baseline transition, under low event rates with a frequent visit schedule (A) or high event rates with a frequent or sparse visit schedule (B and C) and two correlation structures among the covariates (Independent or group-Toeplitz).}
    \label{MSEP_weib}
\end{figure}
\end{document}